# Van der Waals Multiferroic Tunnel Junctions

*Yurong Su,*[†] *Xinlu Li,*[§] *Meng Zhu,*[§] *Jia Zhang,*[*,§] *Long You,*[*,†] *and Evgeny Y. Tsymbal*[*,‡]

[†]School of Optical and Electronic Information, Huazhong University of Science and Technology, 430074 Wuhan, China

[§]School of Physics and Wuhan National High Magnetic Field Center, Huazhong University of Science and Technology, 430074 Wuhan, China

[‡]Department of Physics and Astronomy & Nebraska Center for Materials and Nanoscience, University of Nebraska, Lincoln, Nebraska 68588, USA

## Abstract

Multiferroic tunnel junctions (MFTJs) have aroused significant interest due to their functional properties useful for non-volatile memory devices. So far, however, all the existing MFTJs have been based on perovskite-oxide heterostructures limited by a relatively high resistance-area (*RA*) product unfavorable for practical applications. Here, using first-principles calculations, we explore spin-dependent transport properties of van der Waals (vdW) MFTJs which consist of two-dimensional (2D) ferromagnetic $Fe_nGeTe_2$ ($n$ = 3, 4, 5) electrodes and 2D ferroelectric $In_2Se_3$ barrier layers. We demonstrate that such $Fe_mGeTe_2/In_2Se_3/Fe_nGeTe_2$ ($m, n$ = 3, 4, 5; $m \neq n$) MFTJs exhibit multiple non-volatile resistance states associated with different polarization orientation of the ferroelectric $In_2Se_3$ layer and magnetization alignment of the two ferromagnetic $Fe_nGeTe_2$ layers. We find a remarkably low *RA* product (around 1 $\Omega\cdot\mu m^2$) which makes the proposed vdW MFTJs superior to the conventional MFTJs in terms of their promise for non-volatile memory applications.

electroresistance, tunneling magnetoresistance, multiple non-volatile resistance states, resistance-area product

A promising spintronic device based on electron tunneling is the magnetic tunnel junction (MTJ),[1] which serves as the key building block in non-volatile magnetic random access memories (MRAMs). Changing the magnetic alignment of two ferromagnetic electrodes in an MTJ from parallel to antiparallel causes a sizable change in tunneling resistance of an MTJ, which is known as the tunneling magnetoresistance (TMR) effect.[2] Functional properties of tunnel junctions can also be enhanced using a ferroelectric barrier.[3] Reversal of the electric polarization of the barrier in a ferroelectric tunnel junction (FTJ) by an applied electric field produces a sizable change in resistance of the junction – the phenomenon known as the tunneling electroresistance (TER) effect.[4][5]

A multiferroic tunnel junction (MFTJ) is a FTJ with ferromagnetic electrodes or equivalently a MTJ with a ferroelectric barrier.[6] In a MFTJ, the TER and TMR effects coexist which makes them interesting both from the fundamental point of view as well as from the point of non-volatile low-power memory device application.[7][8][9] So far, all the considered MFTJs have been based on oxide-perovskite materials which can be epitaxially grown using modern thin-film deposition techniques[10][11][12][13][14][15]. However, the resistance-area (*RA*) product of the conventional MFTJs is typically rather large, ranging from $k\Omega\cdot\mu m^2$ [10][11] to $M\Omega\cdot\mu m^2$ [12][13][14][15], which limits their application in practical devices. For instance, for a recording density of around 200 $Gbit/in^2$, an *RA* product of an MTJ read head should lie below 1 $\Omega\cdot\mu m^2$.[16] Similarly, in high-density MRAMs of about 5 $Gbit/in^2$, the impedance matching condition requires an MTJ cell to have an *RA* product of less than 6 $\Omega\cdot\mu m^2$.[17]

The issue of a large *RA* product for conventional MFTJs mainly stems from the intrinsic properties of ferroelectric materials which has a critical thickness of about few nanometers and a relatively large energy bandgap of about several electron-volts (eV). This critical problem, impeding device application of MFTJs, can be solved using recently discovered two-dimensional (2D) van der Waals (vdW) materials.

The discovery of ferromagnetic and ferroelectric 2D vdW materials offers a new platform for exploring new physical phenomena and potential device applications.[18][19]

Heterostructures based on magnetic vdW materials have revealed novel functionalities. Specifically, it has been reported that the graphite/$CrI_3$/graphite vdW tunnel junctions exhibit a huge magnetoresistance effect over thousands of percent at low temperature.[20][21][22][23] Among known 2D ferromagnetic vdW metals, $Fe_nGeTe_2$ ($n$ = 3, 4, 5) compounds [24][25][26][27][28] exhibit high Curie temperature $T_C$ (about 220 K for $Fe_3GeTe_2$, 280 K for $Fe_4GeTe_2$, and over 300 K for $Fe_5GeTe_2$) and thus are promising for spintronic applications. MTJs based on these ferromagnetic vdW materials have been explored both experimentally [29] and theoretically [30]. In addition, thanks to the weak vdW interactions, the Fermi level pinning which may cause the deterioration of TMR in conventional MTJs is largely avoided in vdW MTJs.[31][32]

In parallel with the discovery of 2D ferromagnetism, 2D ferroelectricity has been predicted and experimentally observed in vdW materials. In particular, a single layer of α-$In_2Se_3$ and a similar class of $III_2$-$VI_3$ vdW materials have been demonstrated to exhibit 2D ferroelectricity with both in-plane and out-of-plane polarization at room temperature.[33][34][35][36] Combining ferromagnetic and ferroelectric materials is interesting for creating functional multiferroic vdW heterostructures with magnetoelectric properties. A recent theoretical work has predicted that reversal of $In_2Se_3$ polarization in a $Cr_2Ge_2Te_6$/$In_2Se_3$ vdW heterostructure could change the magnetic anisotropy of $Cr_2Ge_2Te_6$.[37]

These results point to a possibility of creating an MFTJ entirely based on the recently discovered 2D vdW ferroic materials. Due to the stable polarization of the vdW ferroelectrics, such as $In_2Se_3$, down to the monolayer limit, these 2D materials can be efficiently used as ultrathin tunnel barriers in MFTJs. It is expected that in this new type of vdW MFTJs, the *RA* product should be much lower than that in conventional perovskite-oxide MFTJs due to the 2D ferroelectricity of the vdW barrier and its narrow energy bandgap. Thus, 2D vdW ferroic materials may provide a new and more advanced material platform for MFTJs.

In this work, using first-principles calculations based on density functional theory, we investigate full vdW MFTJs and predict that they constitute reliable functional devices with multiple resistance states and a low *RA* product. As a representative example, we consider MFTJs composed of vdW ferromagnetic metals $Fe_nGeTe_2$ and vdW ferroelectric barrier $\alpha$-$In_2Se_3$, and explore their spin-dependent transport properties depending on the orientation of electric polarization of $In_2Se_3$ and the relative magnetization alignment of $Fe_nGeTe_2$. Without loss of generality, we assume that non-magnetic metal electrodes are made of the 2H phase of $PtTe_2$ (2H-$PtTe_2$). Thus, the overall atomic structure of the considered MFTJs is $PtTe_2/Fe_mGeTe_2/\alpha$-$In_2Se_3/Fe_nGeTe_2/PtTe_2$. Figure 1(a) shows the atomic structure of the $PtTe_2/Fe_4GeTe_2/\alpha$-$In_2Se_3/Fe_3GeTe_2/PtTe_2$ MFTJ for the two opposite ferroelectric polarization orientations of $In_2Se_3$ and two magnetization alignments of $Fe_4GeTe_2$ and $Fe_3GeTe_2$. To calculate the transmission across the MFTJs, we proceed as follows.

The crystal structures of bulk $Fe_nGeTe_2$ ($n$ = 3, 4, 5) are shown in Figure 1(b) and the experimental lattice constants of $Fe_3GeTe_2$, $Fe_4GeTe_2$ and α-$In_2Se_3$ are listed in Supplementary Table S1. The crystal structure of bulk $Fe_{5-x}GeTe_2$ has been experimentally reported to belong either to the *R-3m*[26][27] or *R3m*[28] space group. Here, for simplicity, we use the theoretically predicted crystal structure of $Fe_5GeTe_2$ that belongs to the *P3m1* space group and has A-A stacking.[25] The in-plane lattice constants of $Fe_nGeTe_2$ and α-$In_2Se_3$ have a mismatch of less than 1%, which allows using an (1×1) in-plane unit cell to model $Fe_mGeTe_2$/α-$In_2Se_3$/$Fe_nGeTe_2$ MFTJs. The calculations are performed using Quantum ESPRESSO [38] within the generalized gradient approximation (GGA) for the exchange correlation potential [39] and the ultrasoft pseudopotential [40][41]. A Monkhorst-Pack k-point mesh of 16×16×16 and plane-wave cutoff 40 Ry are used for the self-consistent electronic structure calculations. The vdW interaction is taken into account using the DFT-D3 scheme.[42] For the interface structures, atomic relaxations are performed using a 16×16×1 Monkhorst-Pack grid for k-point sampling, and atomic positions are converged until the Hellmann-Feynman forces on each atom become less than $10^{-4}$ Ry/a.u. (~ 2.6 meV/Å).

First, we calculate the electronic structures of bulk $Fe_nGeTe_2$ and one quintuple layer (QL) thickness of $In_2Se_3$. Supplementary Figure S1 shows the calculated spin-polarized band structure and density of states (DOS) of bulk $Fe_nGeTe_2$ which we find consistent with the previous results [25]. The calculated average magnetic moment on Fe atoms is 2.10 $\mu_B$, 2.15 $\mu_B$ and 2.11 $\mu_B$ for $n$ = 3, 4, 5 respectively, which are somewhat larger than the experimental values for $Fe_3GeTe_2$ (1.63 $\mu_B$) [24], $Fe_4GeTe_2$ (1.8 $\mu_B$) [25] and $Fe_5GeTe_2$ (2.0 $\mu_B$) [27]. This discrepancy may partly be attributed to the imperfect stoichiometry in the experimentally prepared samples.

Supplementary Figure S2 shows the calculated band structure of 1QL and 3QL $\alpha$-$In_2Se_3$, indicating that 1QL $\alpha$-$In_2Se_3$ is a semiconductor with an indirect bandgap of 0.80 eV in agreement with the previous results (0.78 eV) [33], and the 3QL $\alpha$-$In_2Se_3$ is a metal with the energy bands crossing the Fermi energy. Due to the presence of a vacuum layer, the out-of-plane ferroelectric polarization of 1QL and 3QL $\alpha$-$In_2Se_3$ is well defined and can be evaluated by directly integrating the charge density (see Supplementary section 2 for calculation details). We find that the out-of-plane electric dipole of 1QL $\alpha$-$In_2Se_3$ (3QL $\alpha$-$In_2Se_3$) is 0.092 eÅ (0.17 eÅ) in good agreement with the previous result of 0.094 eÅ [33]. The in-plane electric dipole of 1QL $\alpha$-$In_2Se_3$ is calculated using the Berry phase method [43] to be 2.694 eÅ which is slightly larger than the value of 2.360 eÅ found in ref. [33], possibly due to the smaller in-plane lattice constant of $In_2Se_3$ ($a$ = 4.026 Å) we have taken from experiments [44] ($a$ = 4.106 Å in ref. [33]).

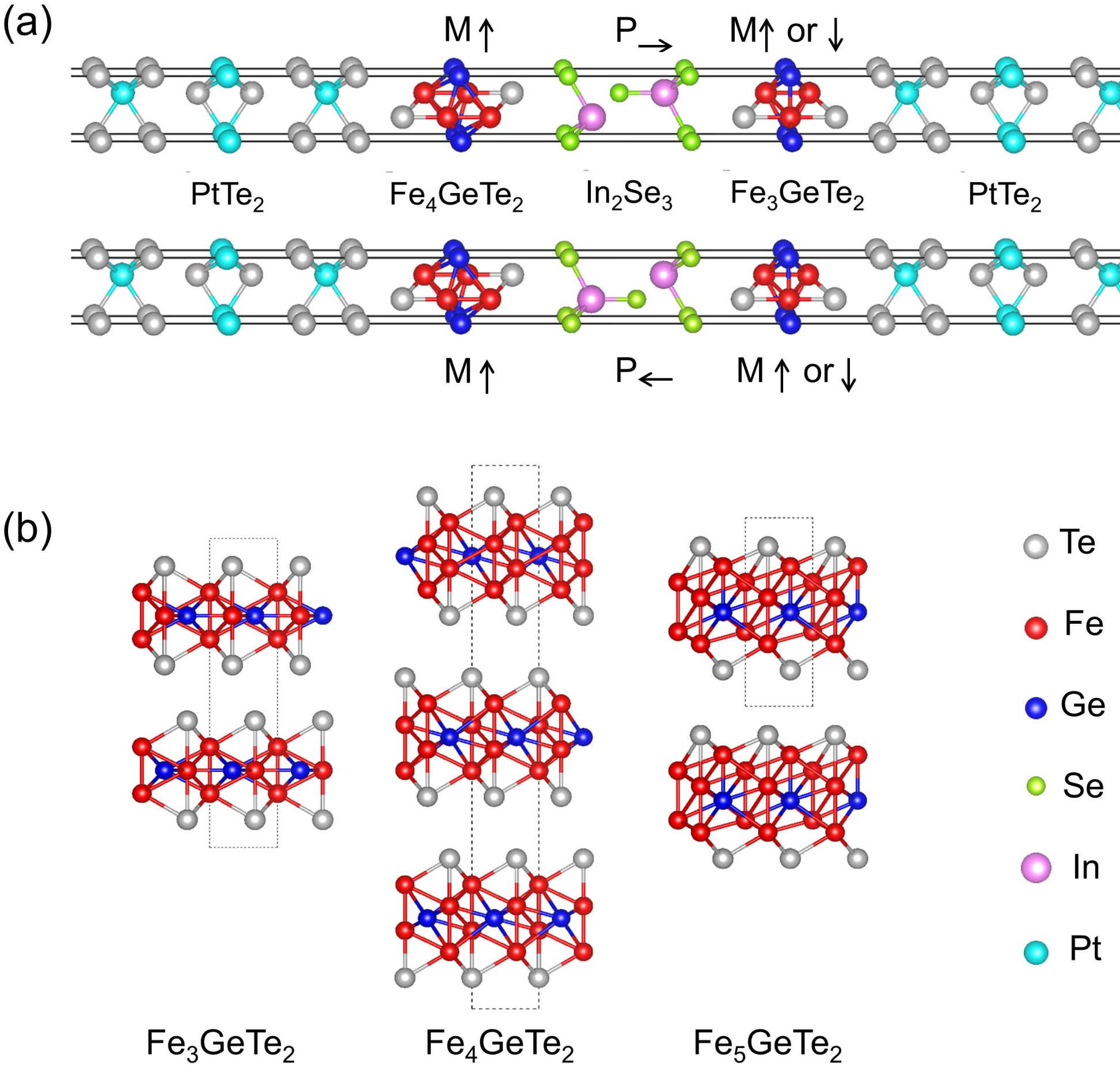

**Figure 1.** (a) Atomic structures of the $PtTe_2/Fe_4GeTe_2/\alpha\text{-}In_2Se_3/Fe_3GeTe_2/PtTe_2$ MFTJs. Ferroelectric polarization of $In_2Se_3$ is pointed right ($P_{\rightarrow}$) or left ($P_{\leftarrow}$). For the calculation of TMR, the magnetic moments of Fe atoms in $Fe_4GeTe_2$ are pointing up ($M_{\uparrow}$), while these in $Fe_3GeTe_2$ are pointing up ($M_{\uparrow}$) or down ($M_{\downarrow}$) for parallel or antiparallel magnetization alignment, respectively. (b) From left to right, the atomic structure of bulk $Fe_3GeTe_2$, $Fe_4GeTe_2$ and $Fe_5GeTe_2$ in a hexagonal lattice.

Next, we consider several possible interface atomic structures. The details of these calculations are given in Supplementary section 3. To build the entire atomic structure of MFTJs with the 2H-$PtTe_2$ electrodes, we fix the in-plane lattice constant to be 4.03 Å for $PtTe_2/Fe_4GeTe_2/\alpha\text{-}In_2Se_3/Fe_3GeTe_2/PtTe_2$ and 4.026 Å for $PtTe_2/Fe_5GeTe_2/\alpha\text{-}In_2Se_3/Fe_3GeTe_2/PtTe_2$ MFTJs. By performing full atomic relaxations of these MFTs, we find the energetically favorable interface between $PtTe_2$ and $Fe_nGeTe_2$ shown in Figure 1(a). Then, for each MFTJ, we calculate self-consistently the electronic structure of the supercell (such as those shown in Figure 1(a)) and the bulk 2H-$PtTe_2$ electrode. The supercell is considered as the scattering region, ideally attached on both sides to semi-infinite 2H-$PtTe_2$ electrodes. Next, the transport properties are calculated using the wave-function scattering method [45] implemented in Quantum ESPRESSO [38]. The spin-dependent conductance of the tunnel junction per unit cell area is calculated as follows: $G_\sigma = \frac{e^2}{h}\sum_{\mathbf{k}_\parallel} T_\sigma(\mathbf{k}_\parallel)$, where $T_\sigma(\mathbf{k}_\parallel)$ is the transmission probability for an electron at the Fermi energy with spin $\sigma$ and Bloch wave vector $\mathbf{k}_\parallel = (k_x, k_y)$, $e$ is the elementary charge, and $h$ is the Plank constant. In the calculations, the 2D Brillouin zone (2DBZ) is sampled using a uniform 200×200 $\mathbf{k}_\parallel$ mesh.

We first investigate the stability of ferroelectric polarization of the 1QL $\alpha$-$In_2Se_3$ sandwiched between the $Fe_nGeTe_2$ layers. Figure 2(a) shows the in-plane averaged macroscopic electrostatic potential across the $Fe_4GeTe_2/\alpha\text{-}In_2Se_3/Fe_3GeTe_2$ junction for the right ($P_\rightarrow$) and left ($P_\leftarrow$) polarization states of $In_2Se_3$. The switchable build-in electric field within the $In_2Se_3$ layer (dashed lines in Figure 2(a)) confirms the presence of ferroelectric polarization in $In_2Se_3$. We find however that the net ferroelectric polarization of the whole $Fe_4GeTe_2/\alpha\text{-}In_2Se_3/Fe_3GeTe_2$ structure is reduced compared to the polarization for a free-standing 1QL $\alpha$-$In_2Se_3$. Specifically, for the $P_\rightarrow$ state, we obtain polarization of 0.022 eÅ, which is around a quarter of the value for a free-standing $\alpha$-$In_2Se_3$ monolayer. This reduction of the ferroelectric polarization can be attributed to the charge transfer between $Fe_nGeTe_2$ and $In_2Se_3$.[46]

An important point to address is the stability of the ferroelectric state of $\alpha$-$In_2Se_3$ in the $Fe_4GeTe_2/\alpha\text{-}In_2Se_3/Fe_3GeTe_2$ structure against transition to the $\beta$'-$In_2Se_3$ phase. We calculate the total energy of the $Fe_4GeTe_2$/1QL-$In_2Se_3/Fe_3GeTe_2$ heterostructure across the ferroelectric phase transition involving both the $\alpha$ and $\beta$' phases of $In_2Se_3$. The results shown in Figure 2(b) demonstrate that both polarization states of $\alpha$-$In_2Se_3$ have a lower total energy compared to the $\beta'$-$In_2Se_3$ phase, which confirms the bi-stable ferroelectric state of $\alpha$-$In_2Se_3$ in the sandwiched structure.

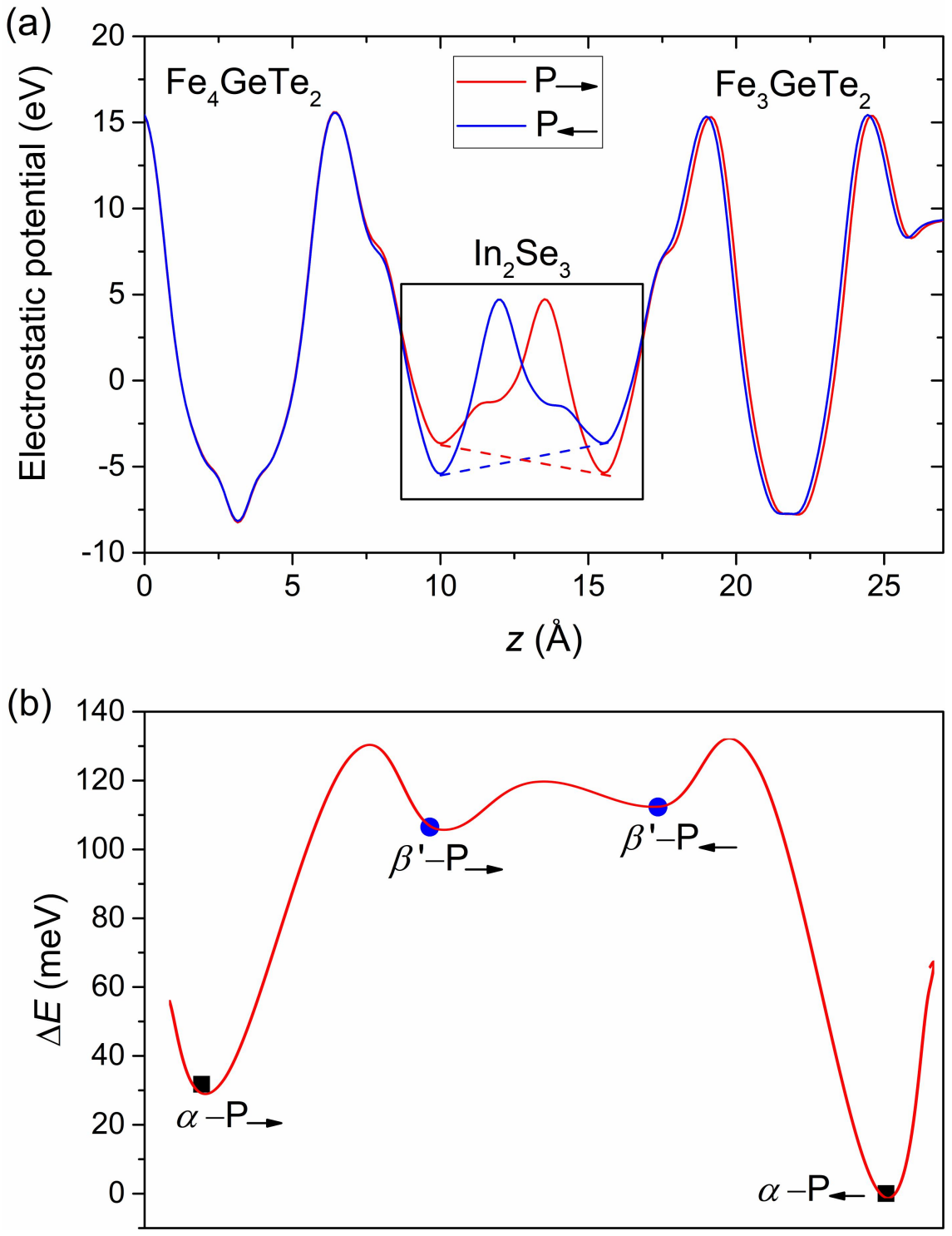

**Figure 2.** (a) The planar averaged macroscopic electrostatic potential in a $Fe_4GeTe_2$/$\alpha$-$In_2Se_3$/$Fe_3GeTe_2$ MFTJ for right ($P_{\rightarrow}$) (red lines) and left ($P_{\leftarrow}$) (blue lines) ferroelectric polarization. The black box indicates the region within 1QL-$In_2Se_3$. The dashed lines in the black box connect the potentials between the two outmost Se layers with a slope indicating the build-in electric field induced by the electric polarization of 1QL-$In_2Se_3$. (b) The total energy profile for a $Fe_4GeTe_2$/1QL-$In_2Se_3$/$Fe_3GeTe_2$ MFTJ across ferroelectric polarization reversal of $\alpha$-$In_2Se_3$ involving ferroelectric states of $\beta'$-$In_2Se_3$.

The calculated transport properties of the $PtTe_2$/$Fe_4GeTe_2$/$\alpha$-$In_2Se_3$/$Fe_3GeTe_2$/$PtTe_2$ MFTJ are shown in Table 1. The TMR ratio is defined as $TMR = (G_P - G_{AP})/G_{AP}$, where $G_P$ and $G_{AP}$ are the conductances for parallel ($M_{\uparrow\uparrow}$) and antiparallel ($M_{\uparrow\downarrow}$) magnetization alignments of the two ferromagnetic $Fe_nGeTe_2$ layers, respectively. Similarly, the TER ratio is defined as $TER = (G_L - G_R)/G_R$, where $G_R$ and $G_L$ are the conductances for right ($P_{\rightarrow}$) and left ($P_{\leftarrow}$) polarization directions of the ferroelectric $In_2Se_3$ layers, respectively. As is evident from Table 1, the TMR depends on ferroelectric polarization of $In_2Se_3$ and is calculated to be 89% and 64%, for right and left polarization orientations, respectively. The TMR originates from the dissimilar electronic structures for majority and minority spins of $Fe_nGeTe_2$ which are evident from the exchange split electronic structures shown in Supplementary Figure S1 as well as the spin-dependent Fermi surfaces shown in Supplementary Figure S6.

**Table 1.** The calculated spin-dependent electron transmission $T_\uparrow$ and $T_\downarrow$, the *RA* product, the TMR (in bold) for left and right polarizations of the $In_2Se_3$ barrier layer and the TER (in bold) for parallel and antiparallel magnetization alignments of the two ferromagnetic $Fe_nGeTe_2$ layers in the $PtTe_2/Fe_4GeTe_2/\alpha$-$In_2Se_3/Fe_3GeTe_2/PtTe_2$ MFTJ.

| | $M_{\uparrow\uparrow}$ (Parallel Magnetization) | | | | $M_{\uparrow\downarrow}$ (Antiparallel Magnetization) | | | | |
|---|---|---|---|---|---|---|---|---|---|
| | Spin up $T_\uparrow$ | Spin down $T_\downarrow$ | *T* ($=T_\uparrow+T_\downarrow$) | *RA* ($\Omega\cdot\mu m^2$) | Spin up $T_\uparrow$ | Spin down $T_\downarrow$ | *T* ($=T_\uparrow+T_\downarrow$) | *RA* ($\Omega\cdot\mu m^2$) | *TMR* |
| $P_\rightarrow$ | $1.48\times10^{-2}$ | $7.11\times10^{-3}$ | $2.19\times10^{-2}$ | 0.17 | $5.82\times10^{-3}$ | $5.76\times10^{-3}$ | $1.16\times10^{-2}$ | 0.31 | **89%** |
| $P_\leftarrow$ | $2.29\times10^{-2}$ | $7.85\times10^{-3}$ | $3.08\times10^{-2}$ | 0.12 | $4.40\times10^{-3}$ | $1.44\times10^{-2}$ | $1.88\times10^{-2}$ | 0.19 | **64%** |
| *TER* | **41%** | | | | **62%** | | | | |

The presence of different ferromagnetic layers, i.e. $Fe_4GeTe_2$ and $Fe_3GeTe_2$, terminating the semi-infinite $PtTe_2$ electrodes produce asymmetry in the MFTJ which is necessary for the non-zero TER effect.[8] We find that the TER ratio is 41% and 62% for parallel and antiparallel magnetization alignment of ferromagnetic layers, respectively (Table 1). These values are not as large as those in some perovskite-oxide FTJs,[10][11] which is explained by the small out-of-plane polarization and the similar atomic and electronic structures of the left and right $Fe_nGeTe_2$ interfaces in the MFTJs.

We find that the ferroelectric polarization reversal of $In_2Se_3$ mainly changes the transmission of the spin-up electrons for parallel magnetization and spin-down electrons for antiparallel magnetization of the ferromagnetic $Fe_4GeTe_2$ and $Fe_3GeTe_2$ layers from Table 1. Noting that the spin-up and spin-down notation is referred to the electron spin in $Fe_4GeTe_2$, this fact indicates that the polarization switching mostly affects the spin-up transmission of $Fe_3GeTe_2$. This can be understood by looking at changes in the interface atomic structure upon ferroelectric polarization switching. Specifically, the interface distance between $Fe_3GeTe_2$ and $\alpha$-$In_2Se_3$ is 0.170 Å for the $P_{\leftarrow}$ state which is smaller than that for the $P_{\rightarrow}$ state (Figure S5). This produces a larger transmission across the interface between the $Fe_3GeTe_2$ and $\alpha$-$In_2Se_3$ layers for the $P_{\leftarrow}$ polarization especially in the spin-up conduction channel, as we will discuss later.

Overall, as seen from Table 1, the electron transmission and the associated *RA* product (see details for the *RA* calculation in Supplementary section 5) reveal four resistance states of the $PtTe_2/Fe_4GeTe_2/\alpha$-$In_2Se_3/Fe_3GeTe_2/PtTe_2$ MFTJ. These states are distinguished by the different magnetic alignments of the ferromagnetic $Fe_nGeTe_2$ layers and different polarization directions of the ferroelectric $In_2Se_3$ layer. It is important to point out that the calculated *RA* products for all four resistance states are less than 1 $\Omega\cdot\mu m^2$ which is a desirable feature of an MFTJ for device applications. This is in contrast to the previously calculated *RA* products for perovskite-oxide MFTJs of around several $k\Omega\cdot\mu m^2$.[10][11] We observe similar features of TMR, TER, and *RA* products for a $PtTe_2/Fe_5GeTe_2/\alpha$-$In_2Se_3/Fe_3GeTe_2/PtTe_2$ MFTJ (see Table S2 in Supplementary Information).

Such low *RA* products in the proposed MFTJ can be attributed to the small bandgap of $\alpha$-$In_2Se_3$. Figure S8 shows the atomic weight projected band structures of the $Fe_4GeTe_2/\alpha$-$In_2Se_3/Fe_3GeTe_2$ MFTJ for ferroelectric polarization pointing right and left. The valence band maximum (VBM) of 1QL $\alpha$-$In_2Se_3$ is found to be 1.14 eV below the Fermi energy due to the band alignment between $Fe_nGeTe_2$ and $In_2Se_3$. As a result, the Fermi energy crosses the conduction bands of $\alpha$-$In_2Se_3$ and leads to the small *RA*.

Increasing $\alpha$-$In_2Se_3$ thickness above 1 QL makes it metallic.[33] Figure S2(b) shows the calculated band structure and DOS of 3QL $\alpha$-$In_2Se_3$ which demonstrate that several energy bands cross the Fermi energy. Table S3 shows the results of transport calculations for a $PtTe_2/Fe_4GeTe_2/\alpha$-$In_2Se_3$ (3QL)/$Fe_3GeTe_2/PtTe_2$ MFTJ where 3QL $\alpha$-$In_2Se_3$ is used as a spacer layer. We find that both TMR and TER are around tens of a percent and the *RA* are lower than 1 $\Omega\cdot\mu m^2$ which prove the robustness of the MFTJ regardless the barrier layer thickness and the conduction type of the $\alpha$-$In_2Se_3$ film.

To elucidate in more detail the effects of ferroelectric polarization and magnetization alignment on electron transmission, we calculate the $\mathbf{k}_{\|}$ - and spin-resolved transmission $T_\sigma(\mathbf{k}_{\|})$ of $PtTe_2/Fe_4GeTe_2/\alpha$-$In_2Se_3/Fe_3GeTe_2/PtTe_2$ MFTJ in the 2DBZ. The results are shown in Figure 3 for parallel ($M_{\uparrow\uparrow}$) and antiparallel ($M_{\uparrow\downarrow}$) magnetization alignments of the $Fe_4GeTe_2$ and $Fe_3GeTe_2$ layers with right ($P_{\rightarrow}$) and left ($P_{\leftarrow}$) ferroelectric polarizations of $In_2Se_3$. The overall transmission patterns reflect the distribution of the available conduction channels of the 2D Fermi surface of the 2H-$PtTe_2$ electrodes displayed in Supplementary Figure S7. There is a relatively large transmission around the $\bar{\Gamma}$ point ($\mathbf{k}_{\|}=0$) of the 2DBZ. It is clearly seen that the transmission of the MFTJ is modulated by reversing the ferroelectric polarization of $In_2Se_3$. Especially, when the ferroelectric polarization is pointing left ($P_{\leftarrow}$), transmission of the spin-up channel for parallel ($M_{\uparrow\uparrow}$) magnetization and transmission of the spin-down channel for antiparallel ($M_{\uparrow\downarrow}$) magnetization are large. That is, the left polarization ($P_{\leftarrow}$) enhances the transmission of the spin-up channel in $Fe_3GeTe_2$ layer.

This behavior can be understood as follows. The transmission for an electron with spin $\sigma$ across a junction can be approximated by the following expression [10][47]:

$$T_\sigma(\mathbf{k}_{\|})=t_L^\sigma(\mathbf{k}_{\|})t_C^\sigma(\mathbf{k}_{\|})t_R^\sigma(\mathbf{k}_{\|})$$

where $t_L^\sigma(\mathbf{k}_{\|})$, $t_R^\sigma(\mathbf{k}_{\|})$ are the interface transmission functions at left ($Fe_4GeTe_2/\alpha$-$In_2Se_3$) and right ($Fe_3GeTe_2/\alpha$-$In_2Se_3$) interfaces, respectively, and $t_C^\sigma(\mathbf{k}_{\|})$ is the transmission function across the $In_2Se_3$ layer. As we discussed previously, for the left ($P_{\leftarrow}$) ferroelectric polarization, the $Fe_3GeTe_2/\alpha$-$In_2Se_3$ interface distance is smaller, and therefore transmission $t_R^\sigma(\mathbf{k}_{\|})$ is larger.

The stronger interface hybridization for the spin-up electrons with the left ($P_{\leftarrow}$) ferroelectric polarization can also be seen from the projected band structure shown in Figure S8.

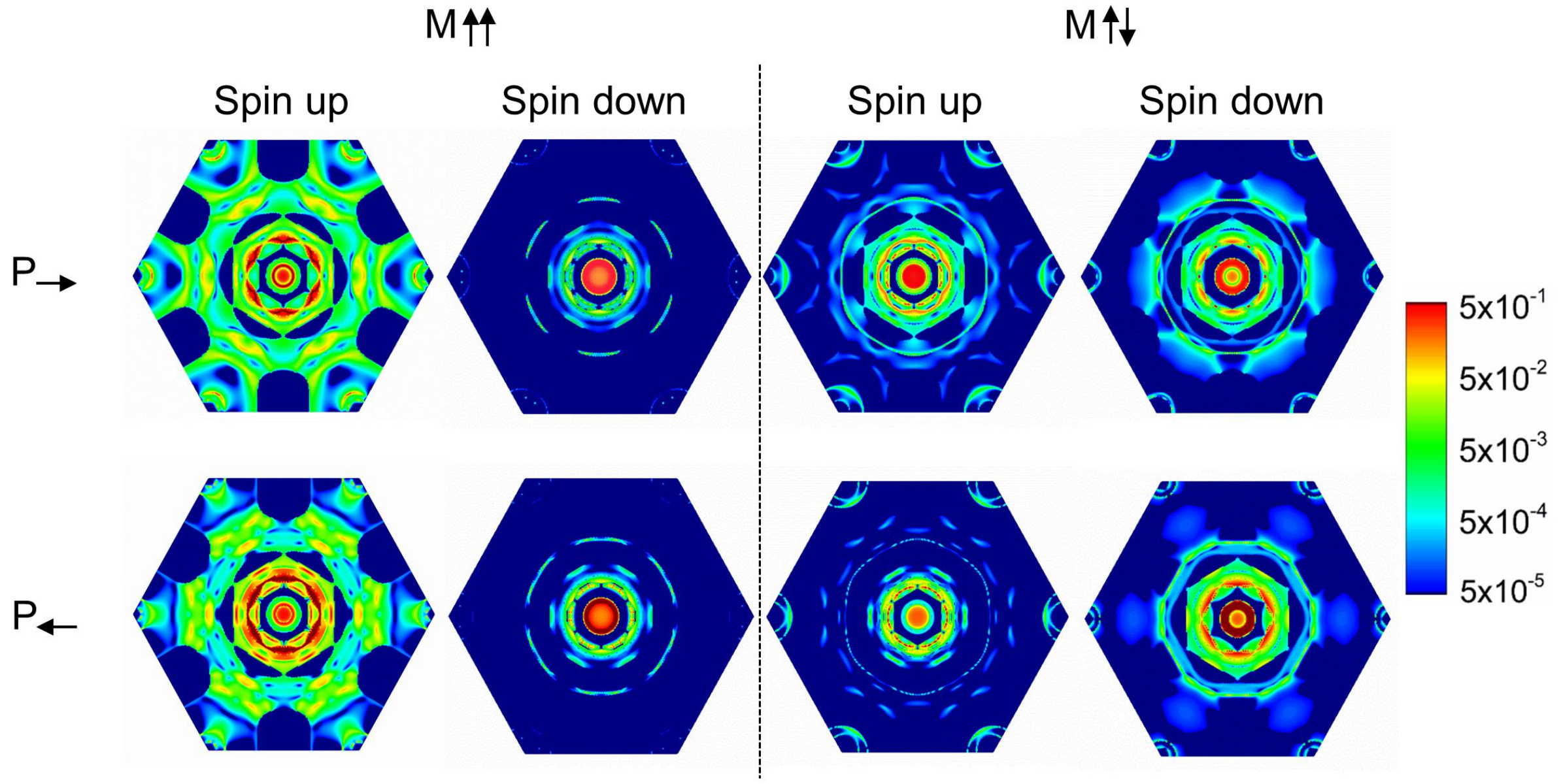

**Figure 3.** The electron transmission for spin-up and spin-down conduction channels in 2DBZ for $PtTe_2/Fe_4GeTe_2/\alpha$-$In_2Se_3/Fe_3GeTe_2/PtTe_2$ MFTJ with right ($P_{\rightarrow}$) and left ($P_{\leftarrow}$) ferroelectric polarizations of $In_2Se_3$ and parallel ($M_{\uparrow\uparrow}$) and antiparallel ($M_{\uparrow\downarrow}$) magnetization alignments of $Fe_4GeTe_2$ and $Fe_3GeTe_2$. The transmission intensity is indicated in logarithmic scale by different colors.

To demonstrate that the multiple resistance states are robust against the electronic structure of the electrodes, we calculate *RA* products for $PtTe_2/Fe_4GeTe_2/In_2Se_3/Fe_3GeTe_2/PtTe_2$ and $PtTe_2/Fe_5GeTe_2/In_2Se_3/Fe_3GeTe_2/PtTe_2$ MFTJs by varying the position of the Fermi energy within the range of $E_F \pm 0.1$ eV. The calculated *RA* products for different polarization and magnetization orientations are shown in Figure 4. One can see that for different Fermi energies, the multiple resistance states are preserved. However, the particular *RA* values of the four resistance states depend on the Fermi energy. For instance, for $PtTe_2/Fe_4GeTe_2/In_2Se_3/Fe_3GeTe_2/PtTe_2$ MFTJ, the multiple resistance states are more separated for $E$ = $E_F$ and $E$ = $E_F$ + 0.1 eV, while for $PtTe_2/Fe_5GeTe_2/In_2Se_3/Fe_3GeTe_2/PtTe_2$ MFTJ, the multiple resistance states are more separated at $E$ = $E_F$ – 0.1 eV. The absolute difference of *RA* does rely on the position of the Fermi energy since at different energies the electronic structures of the electrode, ferromagnetic layers and the barrier are all altered.

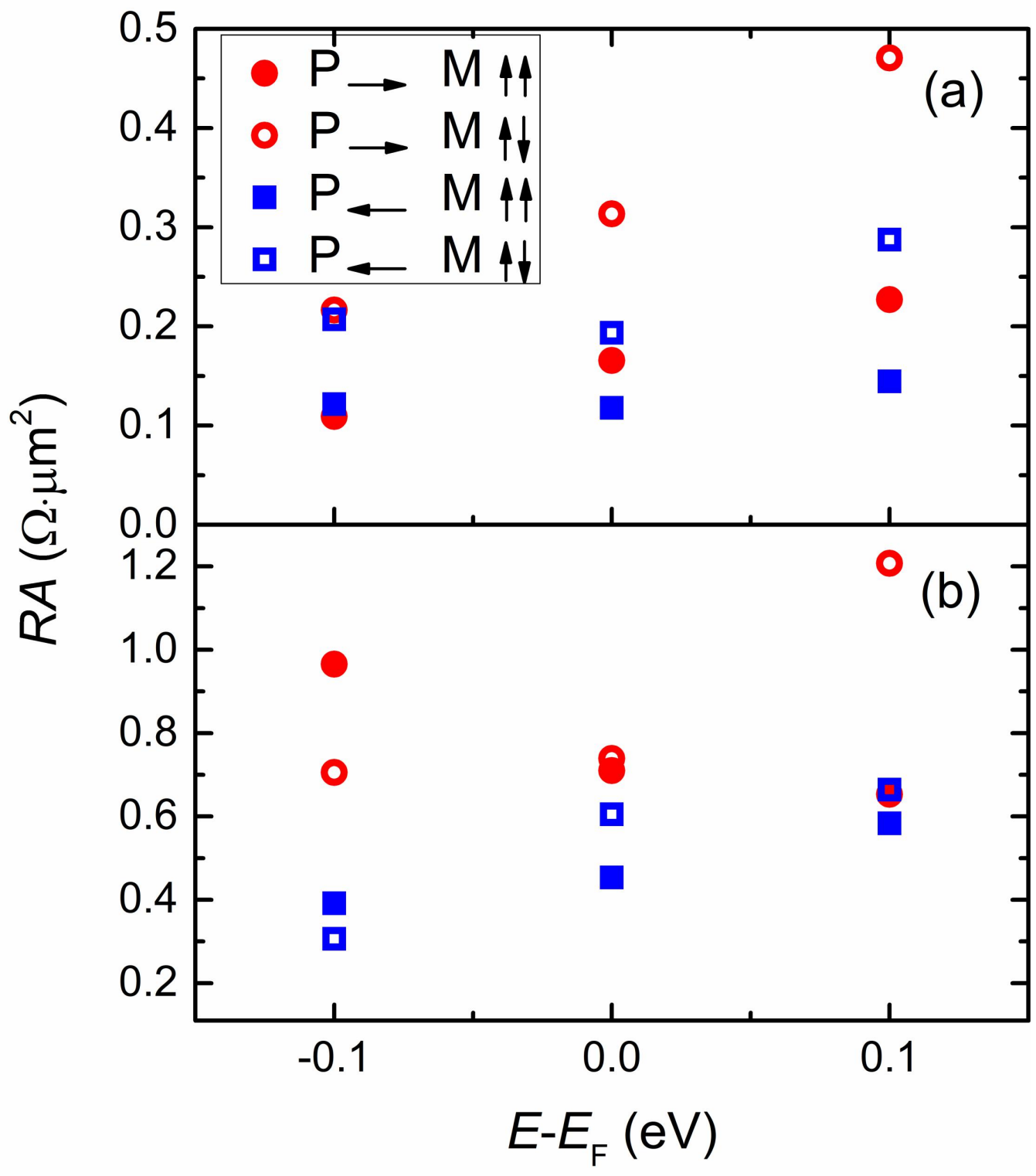

**Figure 4.** *RA* products of the multiple resistance states as a function of the Fermi energy ranging from $E_F$ – 0.1 eV to $E_F$ + 0.1 eV for (a) $PtTe_2/Fe_4GeTe_2/\alpha\text{-}In_2Se_3/Fe_3GeTe_2/PtTe_2$ and (b) $PtTe_2/Fe_5GeTe_2/\alpha\text{-}In_2Se_3/Fe_3GeTe_2/PtTe_2$ MFTJs. The data shown in red circles and blue squares correspond to the right ($P_{\rightarrow}$) and left ($P_{\leftarrow}$) ferroelectric polarizations of $\alpha$-$In_2Se_3$, respectively. The data shown in solid and hollow symbols correspond to the parallel ($M_{\uparrow\uparrow}$) and antiparallel ($M_{\uparrow\downarrow}$) magnetization alignments of the $Fe_nGeTe_2$ layers, respectively.

The proposed MFTJs are feasible for experimental fabrication. Parallel ($M_{\uparrow\uparrow}$) and antiparallel ($M_{\uparrow\downarrow}$) magnetization alignments of $Fe_nGeTe_2$ can be realized thanks to their different coercivities [6] and a weak interlayer exchange coupling (IEC) of two $Fe_nGeTe_2$ layers through $In_2Se_3$ (see section 7 of supporting information). $In_2Se_3$ and similar $III_2$-$VI_3$ vdW ferroelectric materials possess both in-plane and out-of-plane polarizations. That polarization switching can be realized by applying the electric field either in plane or out of plane [36]. Experimentally, out-of-plane (in-plane) switching of $In_2Se_3$ has been achieved in an electric field of 200 kV/cm (40 kV/cm) [48]. Therefore, the non-volatile multiple states with switchable ferroelectric polarizations of $In_2Se_3$ and magnetization alignments of $Fe_nGeTe_2$ can be realized.

The stability of ferroelectric polarization of $In_2Se_3$ QLs sandwiched between magnetic vdW metal layers is an important advantage of vdW ferroelectrics. The low *RA* in the proposed MFTJs stems from the small bandgap of 1QL-$In_2Se_3$ and its conduction bands align with the Fermi level of $Fe_nGeTe_2$, or from the metallic feature when the thickness of $In_2Se_3$ is larger than 2QLs. The TMR and TER in the proposed vdW MFTJs are around tens of a percent, and can be further enhanced by using appropriate materials. For instance, TMR can be enhanced by using thicker $Fe_nGeTe_2$ layers which are expected to recover the spin-polarized bulk band structure [49]. A larger TER can be achieved by exploiting vdW materials with a moderate out-of-plane polarization, or by asymmetric interface engineering. A few examples of this approach are reported in section 8 of supporting information.

In summary, we have introduced a concept of the van der Waals multiferroic tunnel junction, which may serve as a feasible analog of the perovskite-oxide MFTJ, but could produce a better performance for device applications, due to their scalability down to the nanometer layer thickness and a low *RA* product. As an example, we have considered vdW MFTJ consisting of $Fe_nGeTe_2$ ferromagnetic electrodes and $In_2Se_3$ tunnel barriers. Using first-principles calculations based on density functional theory, we have predicted the presence of four resistance states in these vdW MFTJs and an ultralow *RA* product. We hope that our theoretical predictions will stimulate experimental efforts to explore new functionalities of the proposed vdW MFTJs.

## ■ ASSOCIATED CONTENT

**Supporting Information**

Atomic and electronic structure of bulk $Fe_nGeTe_2$ ($n$ = 3, 4, 5); atomic and electronic structure of α-$In_2Se_3$; the calculation of the out-of-plane ferroelectric polarization for 1QL and 3QL α-$In_2Se_3$; interface atomic structure of $Fe_mGeTe_2$/α-$In_2Se_3$/$Fe_3GeTe_2$; electronic structure of the $PtTe_2$/$Fe_mGeTe_2$/α-$In_2Se_3$/$Fe_3GeTe_2$/$PtTe_2$ MFTJ; calculation of the *RA* product; calculated transport properties of the $PtTe_2$/$Fe_5GeTe_2$/α-$In_2Se_3$ (1QL)/$Fe_3GeTe_2$/$PtTe_2$ and the $PtTe_2$/$Fe_4GeTe_2$/α-$In_2Se_3$ (3QL)/$Fe_3GeTe_2$/$PtTe_2$ MFTJs; interlayer exchange coupling of $Fe_nGeTe_2$ through 1QL α-$In_2Se_3$; transport in MFTJs with thicker $Fe_nGeTe_2$ layers and vdW magnet $Co_4GeTe_2$.(PDF)

## ■ AUTHOR INFORMATION

**Corresponding Authors**

*(J. Zhang) E-mail: jiazhang@hust.edu.cn

*(L. You) E-mail: lyou@hust.edu.cn

*(E. Y. Tsymbal) E-mail: tsymbal@unl.edu

**Notes**

The authors declare no competing financial interest.

## ■ ACKNOWLEDGMENTS

Long You and Jia Zhang acknowledge support from the National Natural Science Foundation of China (grants Nos. 11704135, 61674062 and 61821003). Computations were partly performed by utilizing TianHe-2 at the National Supercomputer Center in Guangzhou, China, and the Platform for Data-Driven Computational Materials Discovery at the Songshan Lake Materials Laboratory, Dongguan, China. The authors thank Prof. Jun Sung Kim from Department of Physics, Pohang University of Science and Technology, Korea for providing the structural parameters of bulk $Fe_4GeTe_2$ and $Fe_5GeTe_2$ and Dr. X. K. Huang from Jingdezhen Ceramic Institute, China for helpful discussions.

## ■ REFERENCES

SUPPLEMENTARY INFORMATION

# Van der Waals Multiferroic Tunnel Junctions

*Yurong Su,† Xinlu Li,§ Meng Zhu,§ Jia Zhang,*,§ Long You,*,† and Evgeny Y. Tsymbal*,‡*

†School of Optical and Electronic Information, Huazhong University of Science and Technology, 430074 Wuhan, China

§School of Physics and Wuhan National High Magnetic Field Center, Huazhong University of Science and Technology, 430074 Wuhan, China

‡Department of Physics and Astronomy & Nebraska Center for Materials and Nanoscience, University of Nebraska, Lincoln, Nebraska 68588, USA

**Corresponding Authors**

*(J. Zhang) E-mail: jiazhang@hust.edu.cn

*(L. You) E-mail: lyou@hust.edu.cn

*(E. Y. Tsymbal) E-mail: tsymbal@unl.edu

## 1. Atomic and electronic structure of bulk $Fe_nGeTe_2$ ($n$ = 3, 4, 5)

The experimental lattice structure of bulk $Fe_nGeTe_2$ ($n$ = 3, 4) and the theoretical structure of bulk $Fe_5GeTe_2$ are listed in Table S1. The conventional hexagonal cell of bulk $Fe_3GeTe_2$ consists of two quintuple layers (QLs) stacking in the order of AB. Bulk $Fe_4GeTe_2$ consists of three septuple layers stacking in the order of ABC. $Fe_5GeTe_2$ is stacking as AA.

**Table S1.** The lattice constants and space group of bulk $Fe_nGeTe_2$ ($n$ = 3, 4, 5)[1] and bulk $\alpha$-$In_2Se_3$[2].

| Materials | *a* (Å) | *c* (Å) | Space group/no. |
|---|---|---|---|
| $Fe_3GeTe_2$ | 3.99 | 16.33 | *P6₃/mmc* |
| $Fe_4GeTe_2$ | 4.03 | 29.08 | *R-3m* |
| $Fe_5GeTe_2$ | 4.026 | 10.02 | *P3m1* |
| $\alpha$-$In_2Se_3$ | 4.026 | 28.750 | *R3m* |

The spin-polarized band structure and density of states (DOS) of bulk $Fe_nGeTe_2$ ($n$ = 3, 4, 5) are shown in Figure S1. The exchange splitting between spin-up and spin-down energy bands is visible from the band structure. The spin-up DOS is larger than the spin-down DOS at the Fermi energy. The calculated band structures and DOS agree with the results reported in the reference [1].

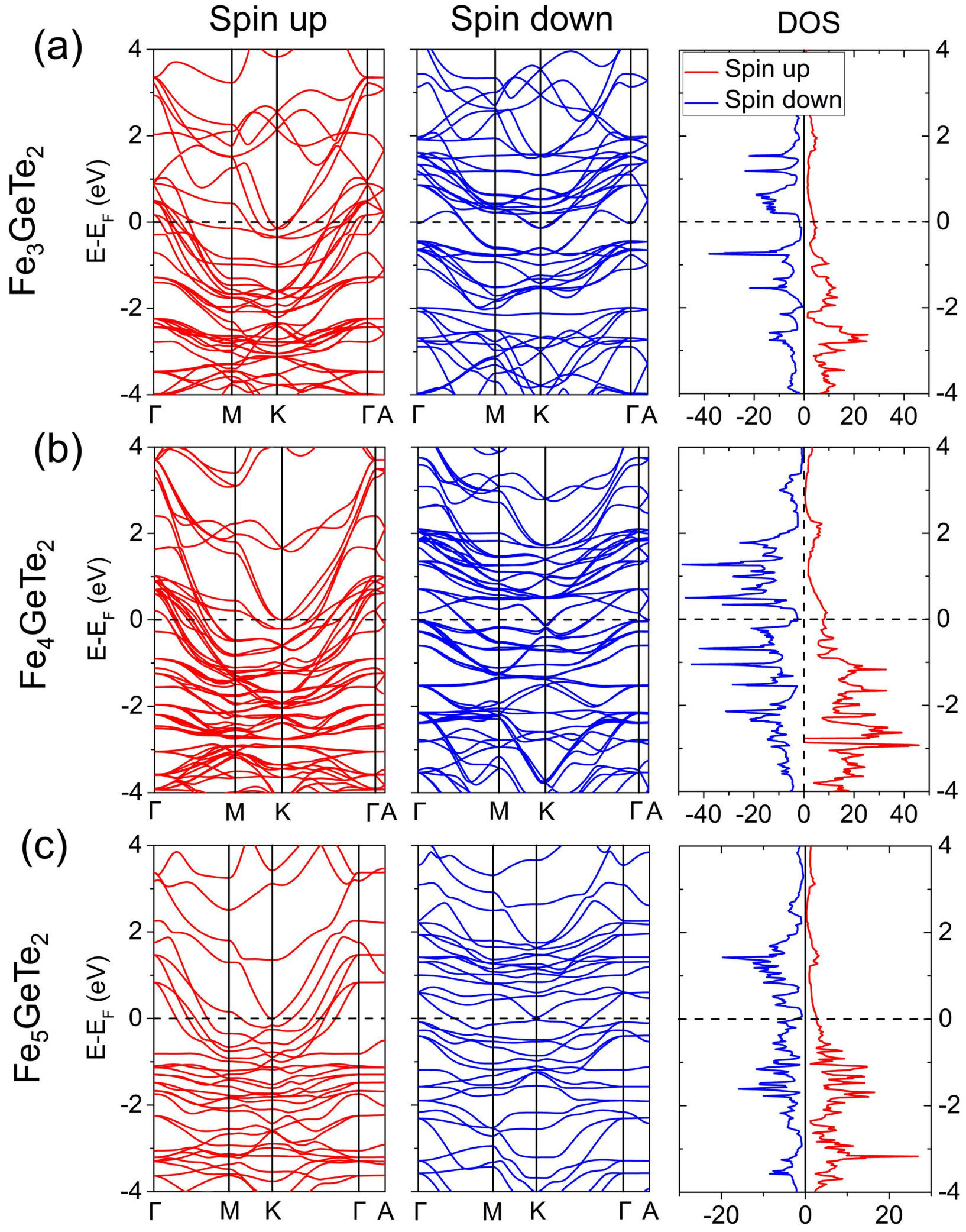

**Figure S1** The calculated band structures and DOS of bulk $Fe_3GeTe_2$ (a), $Fe_4GeTe_2$ (b) and $Fe_5GeTe_2$ (c) in hexagonal lattice. The spin-up and spin-down energy bands and DOS are shown in red and blue,

respectively. The horizontal black dashed lines indicate the position of the Fermi energy.

## 2. Atomic and electronic structure of $\alpha$-$In_2Se_3$

Figure S2 shows the atomic structure, band structure, and DOS of 1QL and 3QL $\alpha$-$In_2Se_3$ with experimental in-plane lattice constant $a$ = 4.026 Å. The calculated energy bands agree with previous results in the reference [2].

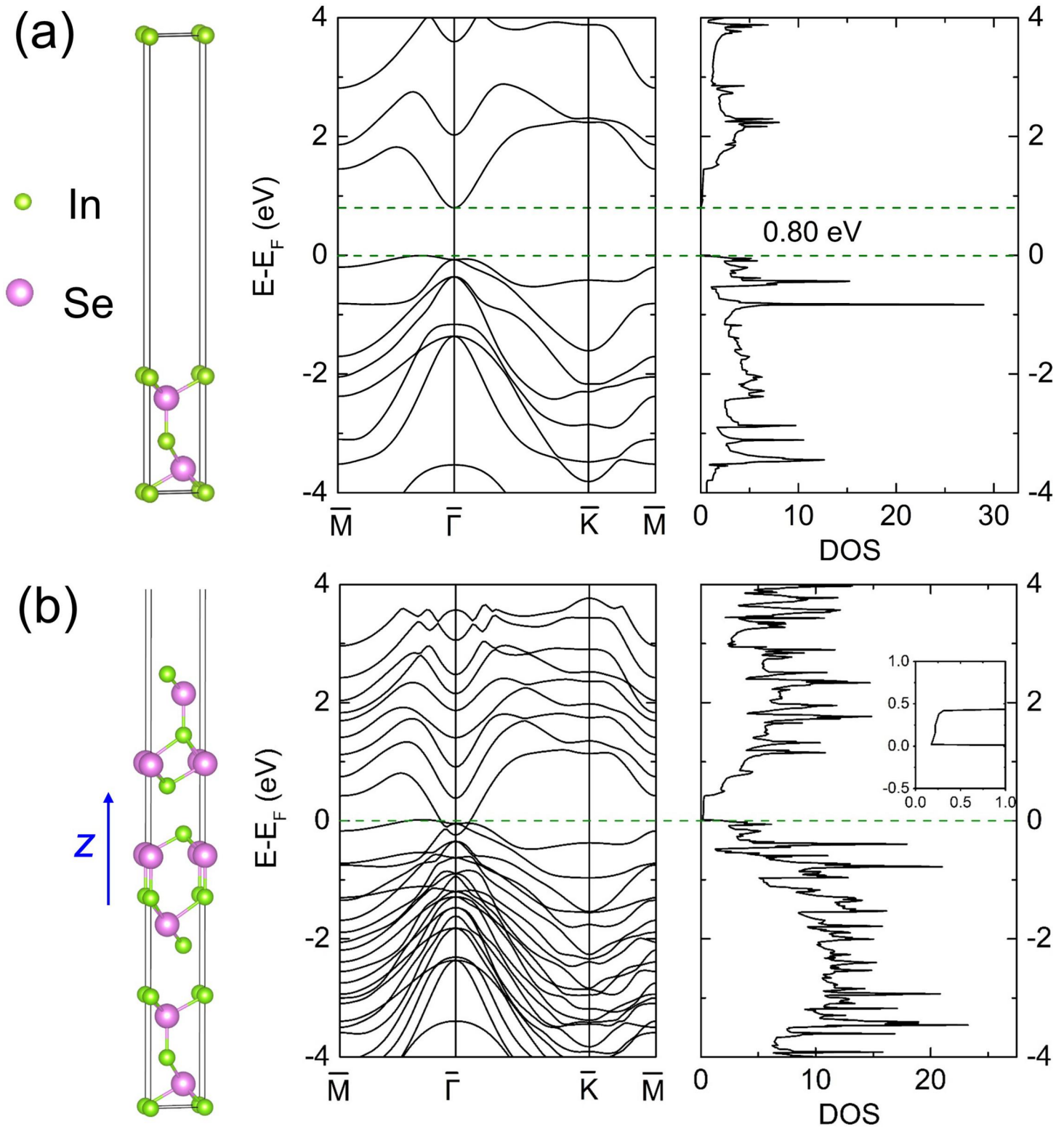

**Figure S2** (a) (Left) The atomic structure of 1QL $\alpha$-$In_2Se_3$ with a vacuum layer of 20 Å. (Right) The band structure and DOS of 1QL $\alpha$-$In_2Se_3$. The two horizontal dashed green lines indicate top of the valence band and bottom of the conduction band. (b) (Left) The atomic structure of 3QL $\alpha$-$In_2Se_3$. The $z$-axis is shown with a blue arrow. (Right) The band structure and DOS of 3QL $\alpha$-$In_2Se_3$. The inset in DOS shows the details around the Fermi energy.

The calculation of the out-of-plane ferroelectric polarization (the electric dipole) for 1QL and 3QL $\alpha$-$In_2Se_3$ is performed using the charge density. The electronic contribution from the valence bands is obtained by integration:

$$P_e = \int_0^c \rho(z) z dz \,,$$

where $\rho(z)$ is the planar-averaged charge density contributed by the valence bands, and $c$ is the cell length along the $z$ direction.

The ionic contribution is obtained by summation over ion cores:

$$P_i = \sum_i Z_i X_i$$

where $i$ is the ion index, $Z_i$ is charge of ion core, and $X_i$ is the ion coordinates along the $z$ direction. The total ferroelectric polarization (the electric dipole) is the sum of these two contributions:

$$P = P_e + P_i \quad .$$

Figure S3 shows the valence electron charge density and the position and charge of the ion cores for 1QL $\alpha$-$In_2Se_3$ and 3QL $\alpha$-$In_2Se_3$.

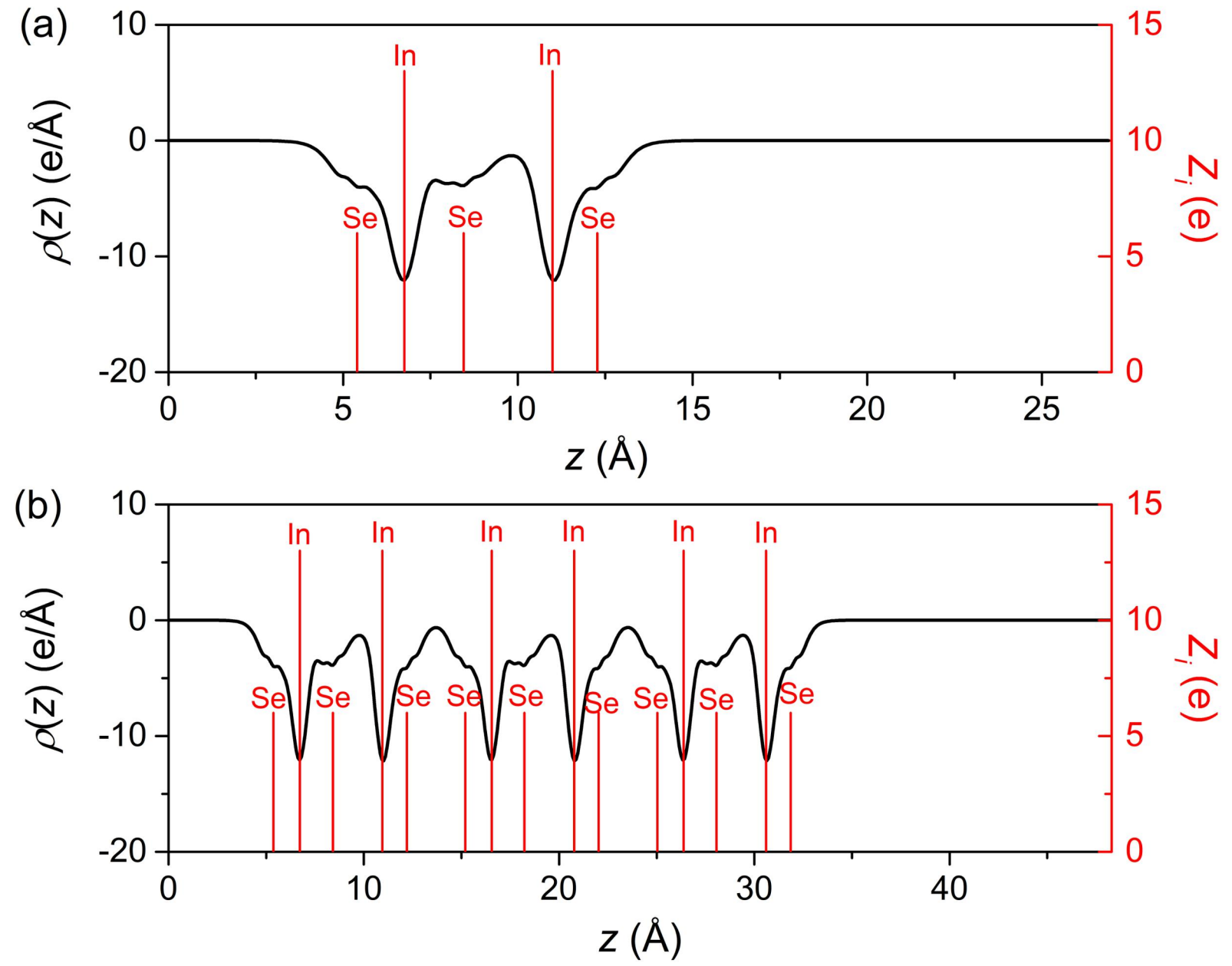

**Figure S3** The planar-averaged valence charge density distribution (black, refers to the left axis) and the charge on the ion cores (red, refers to the right axis) of 1QL $\alpha$-$In_2Se_3$ (a) and 3QL $\alpha$-$In_2Se_3$ (b).

### 3. Interface atomic structure of $Fe_mGeTe_2/\alpha$-$In_2Se_3/Fe_3GeTe_2$

The energetically favorable interface atomic structures between $Fe_3GeTe_2$ and $\alpha$-$In_2Se_3$ have been extensively studied in the reference [3]. We use these interface structures for our $Fe_4GeTe_2/\alpha$-$In_2Se_3/Fe_3GeTe_2$ and $Fe_5GeTe_2/\alpha$-$In_2Se_3/Fe_3GeTe_2$ MFTJs. Below we consider the energetics of the $Fe_4GeTe_2/\alpha$-$In_2Se_3$ and $Fe_5GeTe_2/\alpha$-$In_2Se_3$ which are required to build $Fe_4GeTe_2/\alpha$-$In_2Se_3/Fe_3GeTe_2$ and $Fe_5GeTe_2/\alpha$-$In_2Se_3/Fe_3GeTe_2$ MFTJs.

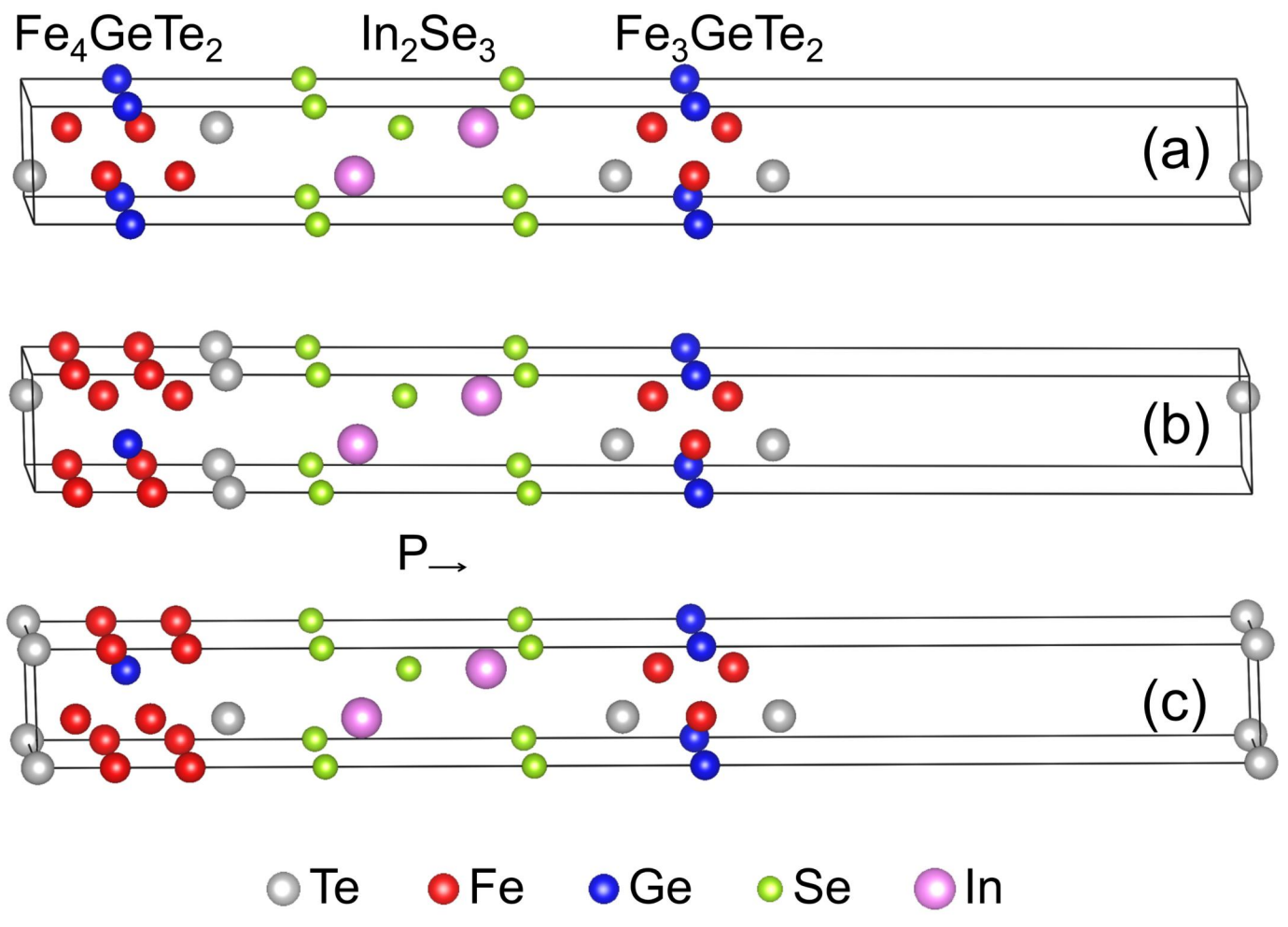

**Figure S4** (a-c) The possible interface atomic structures of $Fe_4GeTe_2/\alpha$-$In_2Se_3/Fe_3GeTe_2$ interface with ferroelectric polarization of $\alpha$-$In_2Se_3$ pointing to right (P→).

For instance, as shown in Figure S4, we consider three possible interface structures between $Fe_4GeTe_2$ and $\alpha$-$In_2Se_3$ in a $Fe_4GeTe_2/\alpha$-$In_2Se_3/Fe_3GeTe_2$ MFTJ. Figure S4(a) shows the interface structure which is found to be most energetically favorable for ferroelectric polarization pointing to right (P→). The total energies of the atomic structures shown in Figures S4(b) and S4(c) are, respectively, 0.27 eV and 0.034 eV higher than that in Figure S4(a). Similarly, we consider the case for ferroelectric polarization pointing to left (P←) and $Fe_5GeTe_2$/$\alpha$-$In_2Se_3$/$Fe_3GeTe_2$ MFTJ for both polarization orientations.

Figure S5 shows the resulting minimum-energy atomic structures for $Fe_4GeTe_2$/$\alpha$-$In_2Se_3$/$Fe_3GeTe_2$ and $Fe_5GeTe_2$/$\alpha$-$In_2Se_3$/$Fe_3GeTe_2$ MFTJs with right (P→) and left (P←) ferroelectric polarizations of $In_2Se_3$. Reversal of ferroelectric polarization alters the interface atomic structures between $Fe_nGeTe_2$ ($n$ = 3, 4, 5) and $In_2Se_3$. For instance, as shown in Figure S5(a), for the $Fe_4GeTe_2$/$\alpha$-$In_2Se_3$/$Fe_3GeTe_2$ interface, polarization switching from right (P→) to left (P←) yields the increase of the $In_2Se_3$-$Fe_3GeTe_2$ interface distance by 0.170 Å and the decrease of the $Fe_4GeTe_2$-$In_2Se_3$ interface distance by -0.044 Å. Similar polarization induced changes in the interface structures occurs for the $Fe_5GeTe_2$/$\alpha$-$In_2Se_3$/$Fe_3GeTe_2$ MFTJ, as shown in Figure S5(b).

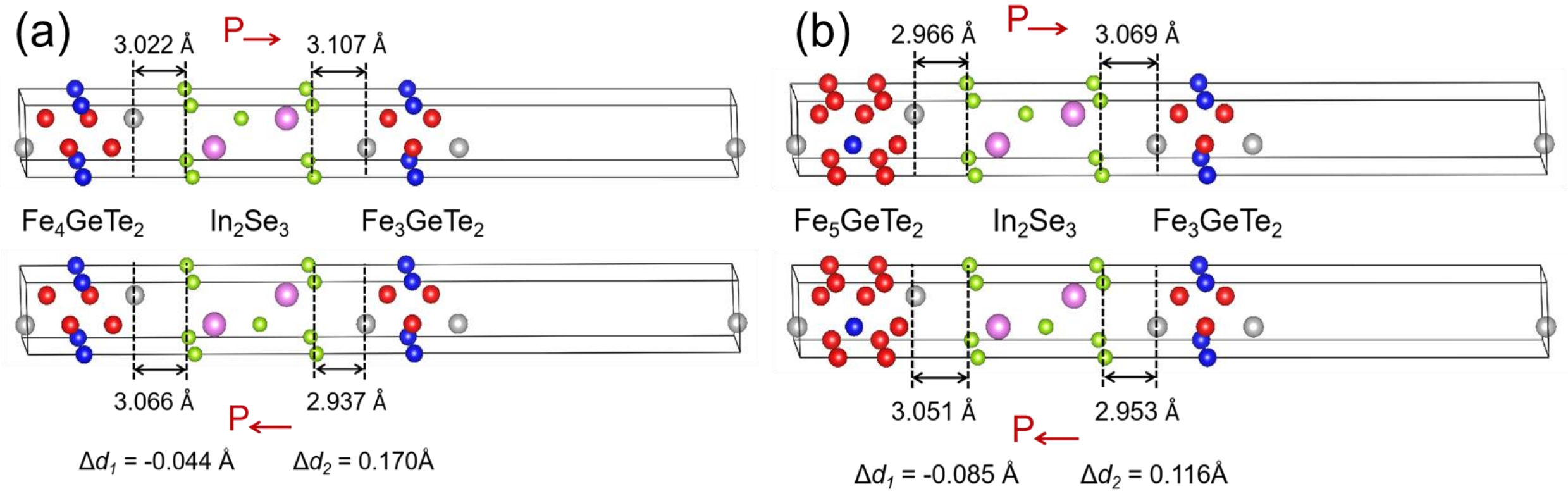

**Figure S5** The energetically optimized interface atomic structure of $Fe_4GeTe_2$/$\alpha$-$In_2Se_3$/$Fe_3GeTe_2$ (a) and $Fe_5GeTe_2$/$\alpha$-$In_2Se_3$/$Fe_3GeTe_2$ (b) with right (P→) and left (P←) ferroelectric polarizations of $In_2Se_3$.

## 4. Electronic structure of the $PtTe_2$/$Fe_mGeTe_2$ (m=4, 5)/$\alpha$-$In_2Se_3$/$Fe_3GeTe_2$/$PtTe_2$ MFTJ

Electronic transport across $PtTe_2$/$Fe_mGeTe_2$ (m=4, 5)/$\alpha$-$In_2Se_3$/$Fe_3GeTe_2$/$PtTe_2$ MFTJ is determined by the available states at the Fermi energy provided by the $PtTe_2$ electrodes and then propagating across the $Fe_mGeTe_2$ layer. The shape of the Fermi surfaces of these materials provides information about the available Bloch states distributed in the Brillouin zone and contributing to the electron transport. Figure S6 shows the calculated spin-polarized Fermi surfaces of bulk $Fe_nGeTe_2$ ($n$ = 3, 4, 5).

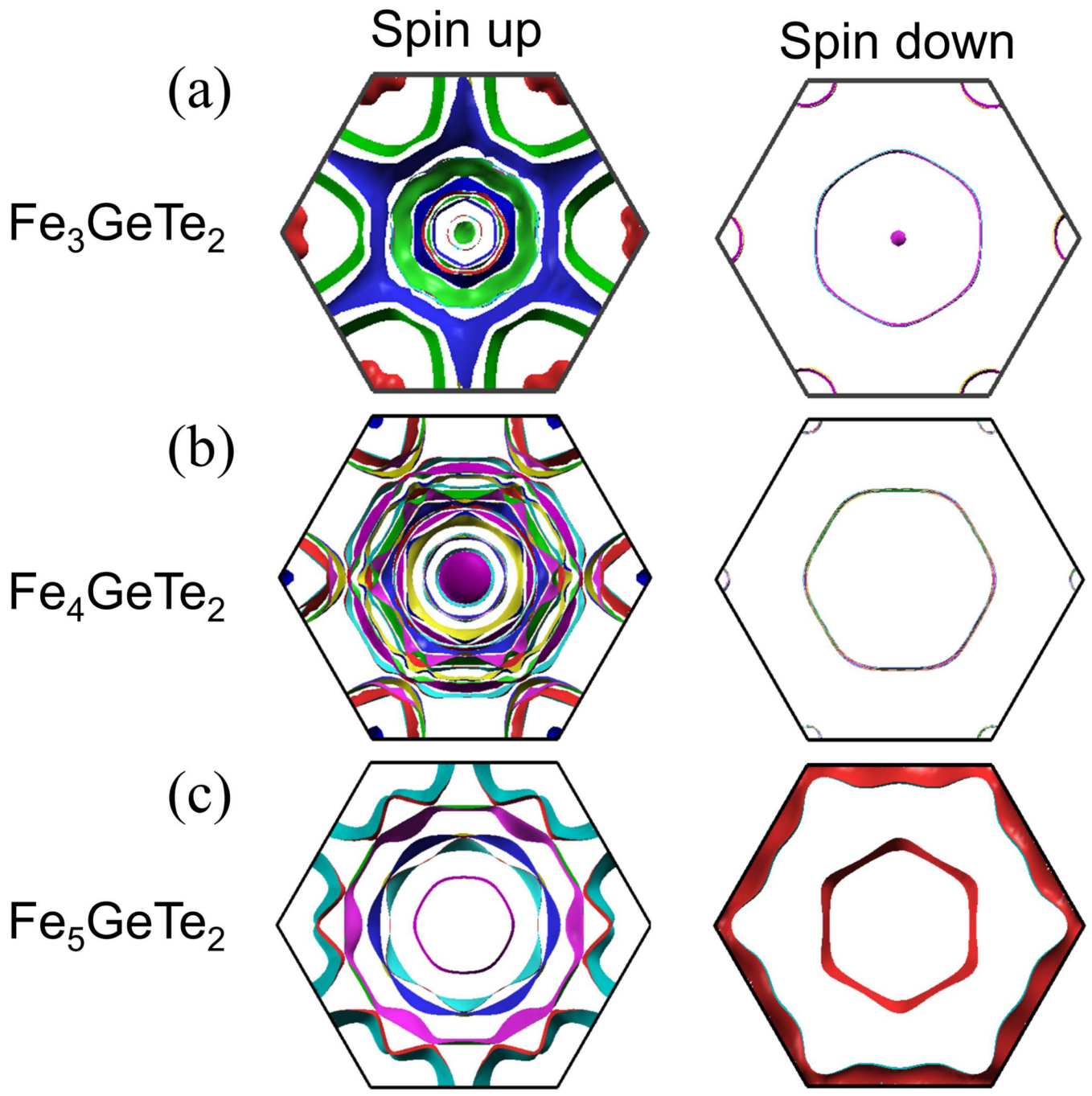

**Figure S6** Spin-resolved Fermi surfaces of bulk $Fe_3GeTe_2$ (a), $Fe_4GeTe_2$ (b) and $Fe_5GeTe_2$ (c).

Semi-infinite 2H-$PtTe_2$ layers are used as metallic electrodes in our MFTJs. Figure S7 (a) shows the unit cell of bulk 2H-$PtTe_2$. The available electronic states for transport across the MFTJ are provided and can be characterized by calculating ballistic transmission across them. For a given wave vector $\mathbf{k}_{\parallel} = (k_x, k_y)$ in the 2D Brillouin zone (2DBZ), the resultant transmission must be integer which reflects the number of available Bloch states at the Fermi energy. Transmission across $PtTe_2/Fe_mGeTe_2/\alpha\text{-}In_2Se_3/Fe_3GeTe_2/PtTe_2$ MFTJs should be consistent with the transmission pattern across $PtTe_2$.

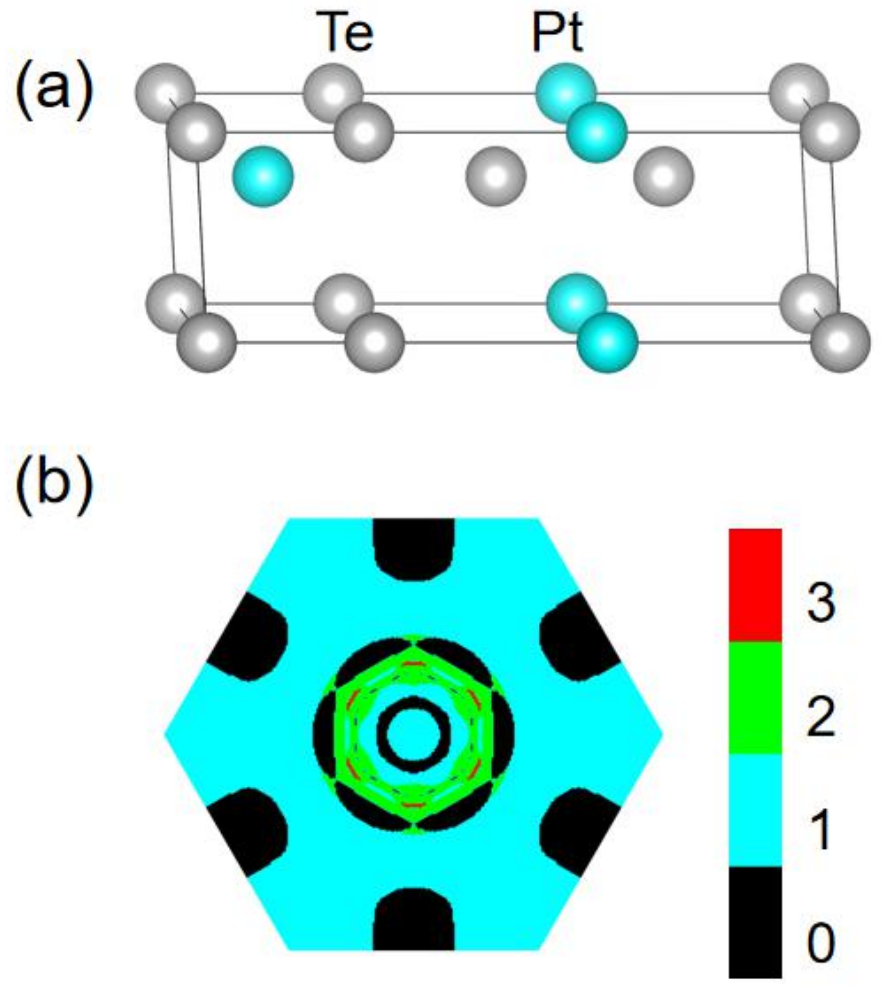

**Figure S7** (a) The atomic structure of bulk 2H-$PtTe_2$. (b) The wave-vector-resolved transmission across bulk 2H-$PtTe_2$ in the 2DBZ. The number at the color scale accounts for the number of available Bloch states.

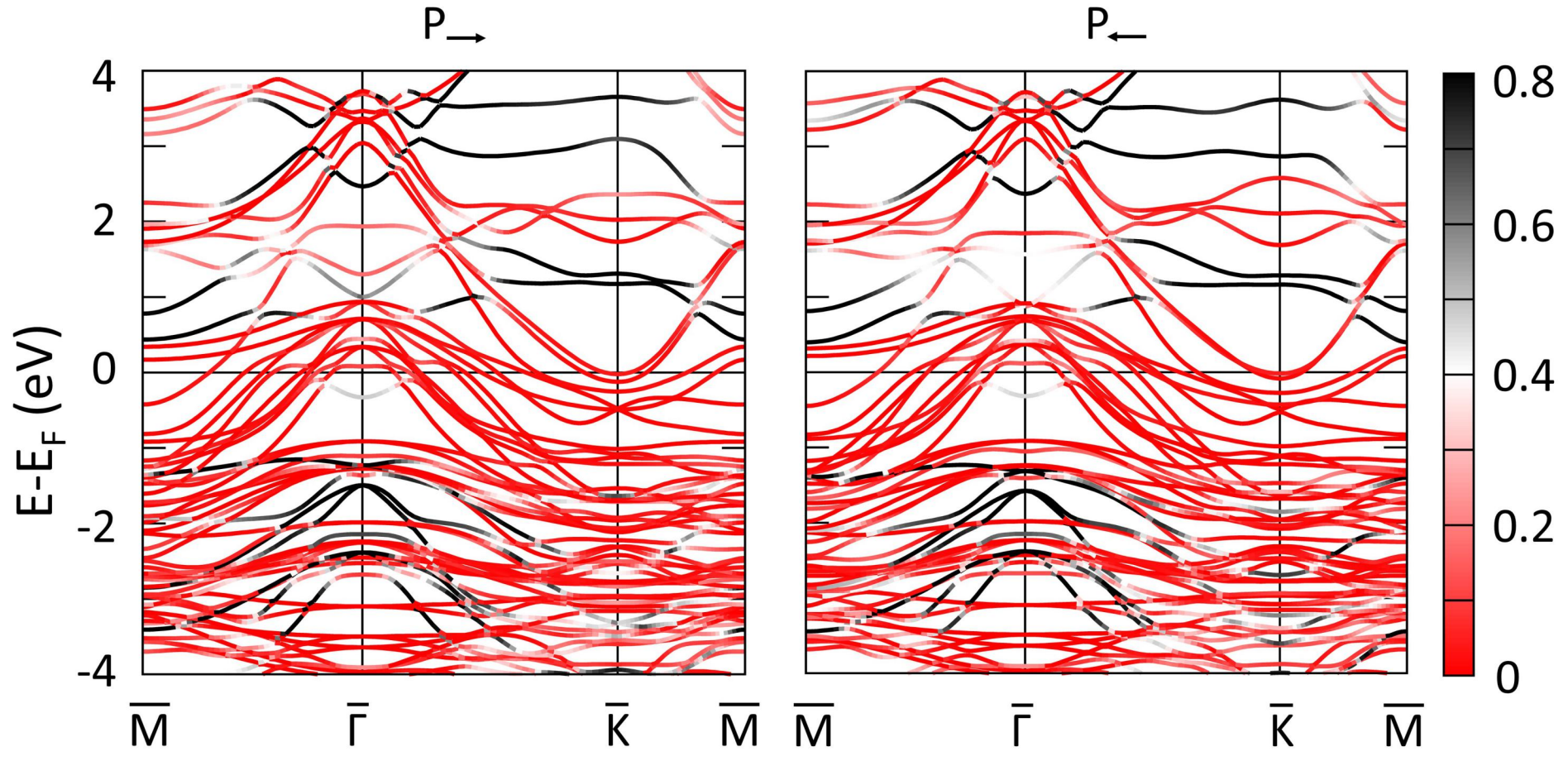

**Figure S8** The atomic weight projected spin-up band structures of the $Fe_4GeTe_2$/$\alpha$-$In_2Se_3$/$Fe_3GeTe_2$ MFTJ with right ($P_{\rightarrow}$) and left ($P_{\leftarrow}$) ferroelectric polarizations in $\alpha$-$In_2Se_3$. The color scale on the right side indicates the portion of the bands projected onto $In_2Se_3$ (black) and $Fe_nGeTe_2$ (red).

## 5. Calculation of the resistance-area product

The resistance-area (RA) product can be calculated from transmission by definition：

$$\text{RA} = \frac{A}{G} = \frac{A}{T(\varepsilon_\text{F})G_0}$$

where $A$ is the unit cell area, $T(\varepsilon_\text{F})$ is the calculated transmission at the Fermi energy, and $G_0 = \frac{e^2}{h} = \frac{1}{25.8k\Omega}$ is the spin-conductance quantum. For instance, in case of $PtTe_2/Fe_4GeTe_2/\alpha\text{-}In_2Se_3/Fe_3GeTe_2/PtTe_2$ MFTJ, the area of the unit cell is: $A = \frac{\sqrt{3}}{2}a^2 = 14.065\,\text{Å}^2$, where $a$ = 4.03 Å is the in-plane lattice of MFTJ. The RA is then given by

$$\text{RA} = \frac{A}{G} = \frac{A}{T(\varepsilon_\text{F})G_0} = \frac{25.8\times14.065\times10^{-5}}{T(\varepsilon_\text{F})}\Omega\cdot\mu m^2 = \frac{3.63\times10^{-3}}{T(\varepsilon_\text{F})}\Omega\cdot\mu m^2$$

## 6. Electron transimission data for $PtTe_2/Fe_5GeTe_2/\alpha\text{-}In_2Se_3$ (1QL)/$Fe_3GeTe_2/PtTe_2$ and $PtTe_2/Fe_4GeTe_2/\alpha\text{-}In_2Se_3$ (3QL)/$Fe_3GeTe_2/PtTe_2$ MFTJs

**Table S2.** The spin resolved electron transmission $T_\uparrow$ and $T_\downarrow$, RA product, TMR (in bold) for right ($P_\rightarrow$) and left ($P_\leftarrow$) polarizations of $In_2Se_3$ and the TER (in bold) for parallel ($M_{\uparrow\uparrow}$) and antiparallel ($M_{\uparrow\downarrow}$) magnetic alignments of two $Fe_nGeTe_2$ layers in $PtTe_2/Fe_5GeTe_2/\alpha\text{-}In_2Se_3$ (1QL)/$Fe_3GeTe_2/PtTe_2$ MFTJ.

| | $M_{\uparrow\uparrow}$ (Parallel Magnetization) | | | | $M_{\uparrow\downarrow}$ (Antiparallel Magnetization) | | | | |
|---|---|---|---|---|---|---|---|---|---|
| | Spin up $T_\uparrow$ | Spin down $T_\downarrow$ | $T$ (=$T_\uparrow$+$T_\downarrow$) | RA ($\Omega\cdot\mu m^2$) | Spin up $T_\uparrow$ | Spin down $T_\downarrow$ | $T$ (=$T_\uparrow$+$T_\downarrow$) | RA ($\Omega\cdot\mu m^2$) | TMR |
| $P_\rightarrow$ | $4.33\times10^{-3}$ | $7.87\times10^{-4}$ | $5.11\times10^{-3}$ | 0.71 | $2.43\times10^{-3}$ | $2.49\times10^{-3}$ | $4.91\times10^{-3}$ | 0.74 | **4%** |
| $P_\leftarrow$ | $7.42\times10^{-3}$ | $5.78\times10^{-4}$ | $8.00\times10^{-3}$ | 0.45 | $1.78\times10^{-3}$ | $4.22\times10^{-3}$ | $6.00\times10^{-3}$ | 0.61 | **33%** |
| TER | **57%** | | | | **22%** | | | | |

**Table S3.** The spin resolved electron transmission $T_\uparrow$ and $T_\downarrow$, RA product, the TM*R* (in bold) for right ($P_\rightarrow$) and left ($P_\leftarrow$) polarizations of $\alpha\text{-}In_2Se_3$ and the TER (in bold) for parallel ($M_{\uparrow\uparrow}$) and antiparallel ($M_{\uparrow\downarrow}$) magnetic alignments of two $Fe_nGeTe_2$ layers in $PtTe_2/Fe_4GeTe_2/\alpha\text{-}In_2Se_3$ (3QL)/$Fe_3GeTe_2/PtTe_2$ MFTJ.

| | $M_{\uparrow\uparrow}$ (Parallel Magnetization) | | | | $M_{\uparrow\downarrow}$ (Antiparallel Magnetization) | | | | |
|---|---|---|---|---|---|---|---|---|---|
| | Spin up $T_{\uparrow}$ | Spin down $T_{\downarrow}$ | $T$ (=$T_{\uparrow}+T_{\downarrow}$) | RA ($\Omega\cdot\mu m^2$) | Spin up $T_{\uparrow}$ | Spin down $T_{\downarrow}$ | $T$ (=$T_{\uparrow}+T_{\downarrow}$) | RA ($\Omega\cdot\mu m^2$) | TMR |
| $P_{\rightarrow}$ | $3.48\times10^{-3}$ | $4.34\times10^{-4}$ | $7.83\times10^{-3}$ | 0.46 | $4.46\times10^{-3}$ | $1.73\times10^{-3}$ | $6.19\times10^{-3}$ | 0.59 | **26%** |
| $P_{\leftarrow}$ | $4.71\times10^{-3}$ | $4.40\times10^{-4}$ | $9.10\times10^{-3}$ | 0.40 | $2.56\times10^{-3}$ | $1.84\times10^{-3}$ | $4.40\times10^{-3}$ | 0.83 | **107%** |
| TER | **26%** | | | | **-29%** | | | | |

## 7. Interlayer exchange coupling of $Fe_nGeTe_2$ through 1QL $\alpha$-$In_2Se_3$

The thickness of 1QL $In_2Se_3$ (including the interface gap) in $Fe_4GeTe_2$/$\alpha$-$In_2Se_3$/$Fe_3GeTe_2$ and $Fe_5GeTe_2$/$\alpha$-$In_2Se_3$/$Fe_3GeTe_2$ is around 1 nm, and consequently may produce sizable interlayer exchange coupling (IEC). The type of IEC indicates the ground magnetization alignments without magnetic field and its magnitude can be defined as:

$$J_{IEC}=E_{AP}-E_{P}$$

where the $E_{AP}$ and $E_P$ are the total energy for parallel and antiparallel magnetic configurations of two $Fe_nGeTe_2$ layers under the same polarization of $In_2Se_3$. By this definition, $J_{IEC} > 0$ ($J_{IEC} < 0$) implies ferromagnetic (antiferromagnetic) IEC, respectively.

Table S4 shows the calculated IEC for $Fe_4GeTe_2$/$\alpha$-$In_2Se_3$/$Fe_3GeTe_2$ and $Fe_5GeTe_2$/$\alpha$-$In_2Se_3$/$Fe_3GeTe_2$ MFTJs with right and left ferroelectric polarizations of $In_2Se_3$. It is seen while the IEC is weak antiferromagnetic for $Fe_4GeTe_2$/$\alpha$-$In_2Se_3$/$Fe_3GeTe_2$, it is ferromagnetic for $Fe_5GeTe_2$/$\alpha$-$In_2Se_3$/$Fe_3GeTe_2$. Reversal of ferroelectric polarization of $In_2Se_3$ alters the magnitude of $J_{IEC}$ slightly but does not change its sign. The IEC for both MFTJs is weak and comparable to that in conventional MgO-based MTJs with ultrathin tunnel barrier[4] which means that non-volatile parallel ($M_{\uparrow\uparrow}$) and antiparallel ($M_{\uparrow\downarrow}$) alignments of $Fe_nGeTe_2$ can be realized by applying magnetic field.

**Table S4.** The calculated IEC (in units of meV/u.c. and mJ/m$^2$) for $Fe_4GeTe_2$/$\alpha$-$In_2Se_3$/$Fe_3GeTe_2$ and $Fe_5GeTe_2$/$\alpha$-$In_2Se_3$/$Fe_3GeTe_2$ MFTJs for right and left ferroelectric polarizations.

| **Interface structures** | **Right ferroelectric polarization ($P_{\rightarrow}$)** | **Left ferroelectric polarization ($P_{\leftarrow}$)** | **Type of IEC** |
|---|---|---|---|
| $Fe_4GeTe_2$/$\alpha$-$In_2Se_3$/$Fe_3GeTe_2$ | $J_{IEC}$ = -0.14 meV/u.c.; -0.16 mJ/m$^2$ | $J_{IEC}$ = -0.036 meV/u.c.; -0.041 mJ/m$^2$ | **AFM** |
| $Fe_5GeTe_2$/$\alpha$-$In_2Se_3$/$Fe_3GeTe_2$ | $J_{IEC}$ = 0.10 meV/u.c.; 0.11 mJ/m$^2$ | $J_{IEC}$ = 0.27 meV/u.c.; 0.31mJ/m$^2$ | **FM** |

## 8. Other vdW MFTJs with enhanced TMR and TER

### 8.1 Transport in the MFTJ with thicker $Fe_nGeTe_2$ layers.

A MFTJ with thicker magnetic layers may have enhanced TMR and TER where the properties of scattering region $Fe_mGeTe_2$/$\alpha$-$In_2Se_3$/$Fe_nGeTe_2$ will play more remarkable role in the transmission of whole MFTJ. To confirm this speculation, we build a MFTJ with the structure of $PtTe_2$/3L-$Fe_4GeTe_2$/$\alpha$-$In_2Se_3$/2L-$Fe_3GeTe_2$/$PtTe_2$, where $Fe_4GeTe_2$ and $Fe_3GeTe_2$ are in three and two layers, respectively. The atomic structure for this MFTJ is shown in Figure S9. The calculated spin-polarized transmission, TMR and TER are summarized in Table S5. The TMR is improved up to hundreds of percent for both polarization orientations. In addition, for AP magnetization state, we found that the conductance for the right polarization ($P_{\rightarrow}$) state is by a factor of 2.8 larger than the conductance for the left polarization ($P_{\leftarrow}$) state, which led to a TER value of 180%. As it is shown in Fig. S10, the polarization modulation of conductance for AP states can be attributed to the transmission around 2DBZ center for spin up channel.

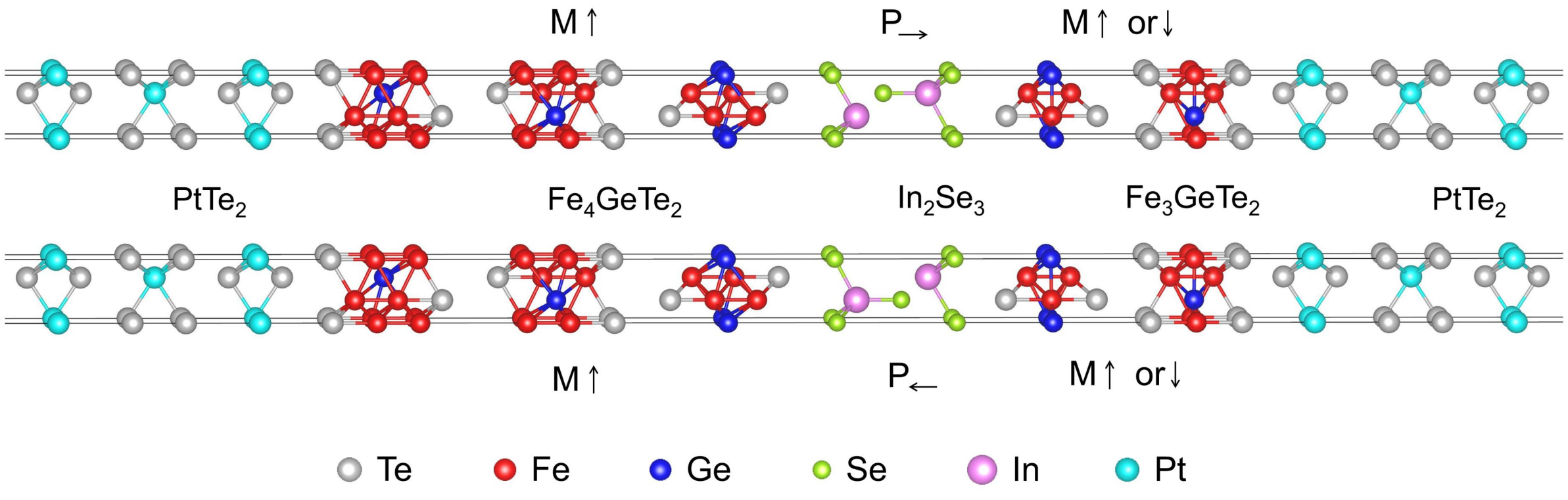

**Figure S9** Atomic structures of the $PtTe_2$/3L-$Fe_4GeTe_2$/$\alpha$-$In_2Se_3$/2L-$Fe_3GeTe_2$/$PtTe_2$ MFTJ for ferroelectric polarization of $In_2Se_3$ pointing right ($P_\rightarrow$) and left ($P_\leftarrow$).

**Table S5.** The calculated spin-dependent electron transmission $T_\uparrow$ and $T_\downarrow$, the RA product, the TMR (in bold) for left and right polarizations of the $In_2Se_3$ barrier layer and the TER (in bold) for parallel and antiparallel magnetization alignments of the two ferromagnets in the $PtTe_2$/3L-$Fe_4GeTe_2$/$\alpha$-$In_2Se_3$/2L-$Fe_3GeTe_2$/$PtTe_2$ MFTJ.

| | $M_{\uparrow\uparrow}$ (Parallel Magnetization) | | | | $M_{\uparrow\downarrow}$ (Antiparallel Magnetization) | | | | |
|---|---|---|---|---|---|---|---|---|---|
| | Spin up $T_\uparrow$ | Spin down $T_\downarrow$ | $T$ (=$T_\uparrow$+$T_\downarrow$) | RA ($\Omega\cdot\mu m^2$) | Spin up $T_\uparrow$ | Spin down $T_\downarrow$ | $T$ (=$T_\uparrow$+$T_\downarrow$) | RA ($\Omega\cdot\mu m^2$) | TMR |
| $P_\rightarrow$ | $1.23\times10^{-2}$ | $4.06\times10^{-5}$ | $1.23\times10^{-2}$ | 0.30 | $4.19\times10^{-3}$ | $3.18\times10^{-4}$ | $4.51\times10^{-3}$ | 0.81 | **173%** |
| $P_\leftarrow$ | $1.82\times10^{-2}$ | $3.16\times10^{-5}$ | $1.82\times10^{-2}$ | 0.20 | $1.06\times10^{-3}$ | $5.50\times10^{-4}$ | $1.61\times10^{-3}$ | 2.25 | **1030%** |
| TER | **50%** | | | | **-64%[a] (-180%[b])** | | | | |

[a]Please note that negative TER indicates larger transmission for right polarization ($P_\rightarrow$).

[b]OTER: The TER according to the optimistic definition: $\mathrm{OTER} = \frac{G_L - G_R}{\min(G_L - G_R)}$, where $G_R$ and $G_L$ are the conductances for right ($P_\rightarrow$) and left ($P_\leftarrow$) polarization directions of the ferroelectric $In_2Se_3$ layers.

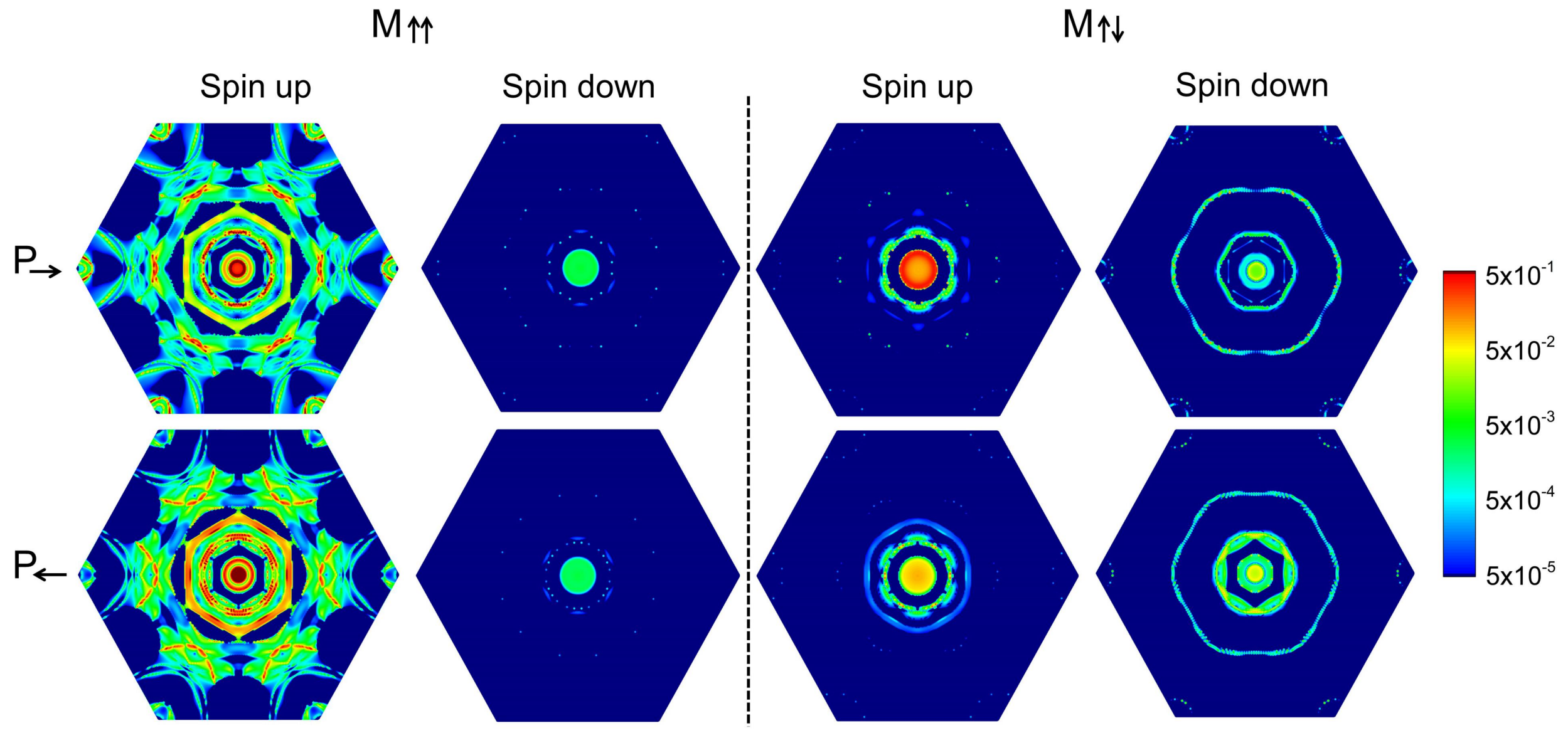

**Figure S10** The electron transmission for spin-up and spin-down conduction channels in 2DBZ for $PtTe_2$/3L-$Fe_4GeTe_2$/$\alpha$-$In_2Se_3$/2L-$Fe_3GeTe_2$/$PtTe_2$ MFTJ with right ($P_{\rightarrow}$) and left ($P_{\leftarrow}$) ferroelectric polarizations of $In_2Se_3$ and parallel ($M_{\uparrow\uparrow}$) and antiparallel ($M_{\uparrow\downarrow}$) magnetization alignments of $Fe_4GeTe_2$ and $Fe_3GeTe_2$. The transmission intensity is indicated in logarithmic scale by different colors.

**8.2 Transport in the MFTJ with vdW magnet of $Co_4GeTe_2$.**

The multiple resistance states in the proposed vdW MFTJs could also be present in the similar structures by using other vdW magnets other that $Fe_nGeTe_2$. To demonstrate the universality of this phenomenon, we setup a MFTJ with structure of $PtTe_2$/$Co_4GeTe_2$/$\alpha$-$In_2Se_3$/$Fe_3GeTe_2$/$PtTe_2$ MFTJ by replacing Fe atoms in one of the $Fe_nGeTe_2$ with cobalt atoms. The atomic structure of this MFTJ is shown in Figure S11. The average magnetic moment per cobalt atom in $Co_4GeTe_2$ is found to be 1.36 $\mu_B$. The calculated transmission, moderate TMR and TER are listed in Table S6. In addition, for parallel magnetic alignments, the transmission for right polarization ($P_{\rightarrow}$) is around 2.2 times larger than the value for left polarization ($P_{\leftarrow}$) state.

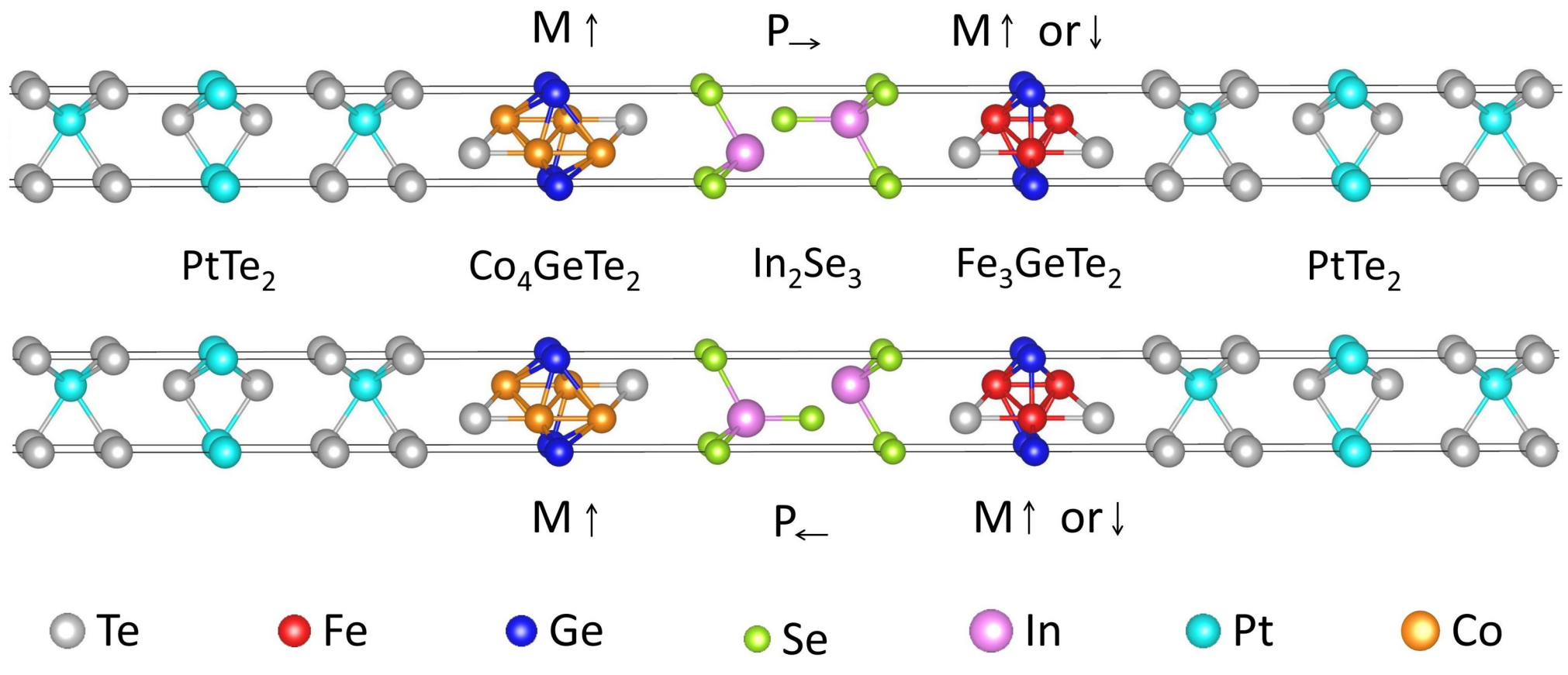

**Figure S11** Atomic structures of the $PtTe_2$/$Co_4GeTe_2$/$\alpha$-$In_2Se_3$/$Fe_3GeTe_2$/$PtTe_2$ MFTJ for ferroelectric polarization of $In_2Se_3$ pointing right ($P_{\rightarrow}$) and left ($P_{\leftarrow}$).

**Table S6.** The calculated spin-dependent electron transmission $T_\uparrow$ and $T_\downarrow$, the RA product, the TMR (in bold) for left and right polarizations of the $In_2Se_3$ barrier layer and the TER (in bold) for parallel and antiparallel magnetization alignments of the two ferromagnets in the $PtTe_2/Co_4GeTe_2/\alpha\text{-}In_2Se_3/Fe_3GeTe_2/PtTe_2$ MFTJ.

| | $M_{\uparrow\uparrow}$ (Parallel Magnetization) | | | | $M_{\uparrow\downarrow}$ (Antiparallel Magnetization) | | | | |
|---|---|---|---|---|---|---|---|---|---|
| | Spin up $T_\uparrow$ | Spin down $T_\downarrow$ | $T$ ($=T_\uparrow+T_\downarrow$) | RA ($\Omega\cdot\mu m^2$) | Spin up $T_\uparrow$ | Spin down $T_\downarrow$ | $T$ ($=T_\uparrow+T_\downarrow$) | RA ($\Omega\cdot\mu m^2$) | TMR |
| $P_\rightarrow$ | $7.82\times10^{-3}$ | $6.94\times10^{-5}$ | $7.89\times10^{-3}$ | 0.46 | $5.98\times10^{-4}$ | $1.42\times10^{-3}$ | $2.02\times10^{-3}$ | 1.80 | **291%** |
| $P_\leftarrow$ | $3.38\times10^{-3}$ | $2.54\times10^{-4}$ | $3.63\times10^{-3}$ | 1.00 | $7.72\times10^{-4}$ | $2.09\times10^{-3}$ | $2.86\times10^{-3}$ | 1.27 | **27%** |
| TER | | **-54% (-117%[a])** | | | | **42%** | | | |

[a]OTER has the same definition as in Table S5.